\documentclass[twocolumn]{aastex631}
\usepackage[version=4]{mhchem}

\usepackage{booktabs, multirow}

\begin{document}

\title[Considerations for Photochemical Modeling of Possible Hycean Worlds]{Considerations for Photochemical Modeling of Possible Hycean Worlds}

\author[0000-0001-6067-0979]{G. J. Cooke}\thanks{E-mail: gjc53@cam.ac.uk},
\affiliation{Institute of Astronomy, University of Cambridge, Cambridge, CB3 0HA, UK.\\}

\author[0000-0002-4869-000X]{N. Madhusudhan}\thanks{E-mail: nmadhu@ast.cam.ac.uk}.
\affiliation{Institute of Astronomy, University of Cambridge, Cambridge, CB3 0HA, UK.\\}

\renewcommand{\thetable}{A\arabic{table}}
\setcounter{table}{0} 

\begin{abstract}
\noindent
JWST is revolutionising the study of temperate sub-Neptunes, starting with the first detection of carbon-bearing molecules in the habitable-zone sub-Neptune K2-18~b. The retrieved abundances of \ce{CH4} and \ce{CO2} and non-detection of \ce{NH3} and \ce{CO} in K2-18~b are consistent with prior predictions of photochemical models for a Hycean world with a habitable ocean. However, recent photochemical modeling raised the prospect that the observed abundances may be explained by a mini-Neptune scenario instead. In this study, we explore these scenarios using independent photochemical modeling with K2-18~b as a case study. We find the previous results to be sensitive to a range of model assumptions, such as the photochemical cross sections, incident stellar spectrum, surface pressure, UV albedo, and metallicity, significantly affecting the resulting abundances. We explore a wide model space to investigate scenarios that are compatible with the retrieved molecular abundances for K2-18~b. Our analysis shows that the previously favoured mini-Neptune scenario is not compatible with most of the retrieved abundances, while the Hycean scenarios, both inhabited and uninhabited, provide better agreement. An uninhabited Hycean scenario explains most of the abundance constraints, except CH$_4$ which is generally underabundant but dependent on the model assumptions. The inhabited Hycean scenario is compatible with all the abundances if the observed CH$_4$ is assumed to be predominantly biogenic. Our results underscore the importance of systematic photochemical modeling and accurate interpretation of chemical abundance constraints for candidate Hycean worlds.
\end{abstract}

\keywords{Exoplanets (498) --- Exoplanet atmospheres (487) --- Habitable planets (695) --- Exoplanet atmospheric composition (2021) --- Biosignatures (2018) --- Mini Neptunes(1063)}

\section{Introduction}
\label{Introduction section}

The search for potentially habitable exoplanets and exploring their diversity is a major goal of modern astronomy. The sub-Neptune regime consists of exoplanets with radii between those of Earth and Neptune ($\sim$1-4 $R_\oplus$). These exoplanets may straddle either side of the radius valley \citep{2017AJ....154..109F, 2018AJ....156..264F} and have been given distinct classifications as super-Earths or mini-Neptunes depending on their size and density \citep{2018AJ....156..264F, 2022Sci...377.1211L, 2024arXiv240504057L}. In the sub-Neptune regime, exoplanets of a certain size can be degenerate between rocky worlds that have large hydrogen-dominated atmospheres, or lower density worlds such as mini-Neptunes or water worlds, with thinner atmospheres \citep{2020ApJ...891L...7M, 2021MNRAS.505.3414N, 2023FaDi..245...80M}.

In the last few years, the possibility of Hycean exoplanets has been proposed \citep{2021ApJ...918....1M}. These planets, with radii between 1.1 -- 2.6 $R_\oplus$ and masses between 1 -- 10 $M_\oplus$, offer unique opportunities for studying their atmospheric and internal structures. Hycean exoplanets are expected to have temperature and pressure conditions that enable a liquid water ocean to exist underneath a hydrogen-rich atmosphere, with a thick layer of ice below a deep ocean \citep{2021ApJ...918....1M, 2024MNRAS.529..409R}. Given their larger radii than rocky exoplanets and the low mean molecular weight of their H$_2$-rich atmospheres, Hycean worlds are favourable targets in the search for habitable environments beyond Earth. Promisingly, the theoretical Hycean habitable zone \citep{2021ApJ...918....1M} is much larger than the purported terrestrial habitable zone \citep{1993Icar..101..108K, 2011AsBio..11..443A, 2013ApJ...765..131K, 2016ApJ...819...84K, 2024arXiv240112204W}. On top of this, Hycean exoplanets could provide the necessary resources for life to survive and flourish, including bio-essential elements and adequate energy sources \citep{2023FaDi..245...80M}.

The James Webb Space Telescope (JWST) has brought in unprecedented capability in the characterisation of temperate sub-Neptunes, demonstrated by the first detection of carbon-based molecules in the candidate Hycean world K2-18~b \citep{2023ApJ...956L..13M}. A transmission spectrum of K2-18~b was observed in the 1-5 $\mu$m range (Cycle 1 GO Program 2722) which revealed a \ce{H2}-dominated atmosphere with abundances of \ce{CH4} and \ce{CO2} of $\sim$1\% by volume  \citep{2023ApJ...956L..13M}. This was the first time carbon-bearing molecules were found in a habitable zone exoplanet. \cite{2023ApJ...956L..13M} did not find evidence for the presence of \ce{H2O} or \ce{NH3}, which could indicate that \ce{NH3} may have been photodissociated and/or dissolved in a liquid water ocean \citep{2021ApJ...921L...8H,2021ApJ...922L..27T,2023FaDi..245...80M}.

The exoplanet K2-18~b provides a valuable opportunity for atmospheric characterisation in the temperate sub-Neptune regime. The planetary parameters are shown in Table~\ref{Planetary properties table}. The planetary bulk parameters place K2-18~b in the sub-Neptune regime with multiple solutions for the possible internal structure, including a mini-Neptune, gas dwarf, or a Hycean world \citep{2020ApJ...891L...7M}. The  instellation received by K2-18~b is very similar to that of Earth, with an  equilibrium temperature ($T_\textrm{eq}$) of $\sim$235-280 K for a Bond albedo between 0-0.5 assuming uniform redistribution. Several studies have used K2-18~b as a prototype to investigate atmospheric chemistry in the temperate sub-Neptune regime under different planetary conditions. Such models have spanned a wide range of conditions, including models with a solid/ocean surface and a thin H$_2$-rich atmosphere \citep{2021ApJ...914...38Y, 2021ApJ...921L...8H,   2021ApJ...922L..27T, 2023FaDi..245...80M} as well as mini-Neptunes and gas dwarfs with deep H$_2$-rich atmospheres that may be physically plausible \citep[e.g.][]{2021ApJ...921...27H, 2024ApJ...963L...7W, rigby2024towards}. 

The JWST observations of K2-18~b are beginning to provide important insights into its possible internal structure and surface conditions \citep{2023ApJ...956L..13M}. The retrieved molecular abundances of \ce{CH4} and \ce{CO2} and non-detections of NH$_3$ and CO are consistent with previous predictions for the presence of an ocean under a thin H$_2$-rich atmosphere \citep{2021ApJ...914...38Y, 2021ApJ...921L...8H,   2021ApJ...922L..27T, 2023FaDi..245...80M}, i.e. a Hycean world \citep{2023ApJ...956L..13M}. Following the JWST observations of K2-18~b, some studies have proposed alternative scenarios to explain the observed spectrum, including a mini-Neptune \citep{2024ApJ...963L...7W} or a putative magma ocean in which \ce{NH3} was suggested to dissolve \citep{2024ApJ...962L...8S}. However, neither of these latter scenarios are able to simultaneously explain all the observed abundances, particularly the non-detections of NH$_3$ and CO abundances coupled with a high CO$_2$ abundance,  as noted by \cite{2024ApJ...964L..19G}. Furthermore, \cite{rigby2024towards} found that the magma ocean model solutions proposed by \cite{2024ApJ...962L...8S} were physically implausible as they  
violated mass balance and bulk density, besides being chemically incompatible with several of the retrieved abundance constraints. Other mini-Neptune scenarios proposed are also inconsistent with the retrieved abundances \citep{2024arXiv240709009H, 2024ApJ...971L..48Y, 2024arXiv240906258L}, albeit with different model assumptions; see e.g. Section~\ref{Mini-Neptune scenario results}. For example, \cite{2024arXiv240906258L} do not consider atmospheric photochemistry, making any comparison with photospheric abundances unreliable. 

\cite{2024ApJ...963L...7W}, hereafter W24, stated that the retrieved composition of K2-18~b is broadly consistent with a mini-Neptune but not with an uninhabited Hycean world, while the inhabited Hycean was considered less plausible. The primary distinction between their uninhabited and inhabited Hycean scenarios was that the former is significantly depleted in CH$_4$ whereas the latter has a biogenic source for CH$_4$. The potential for a biogenic source of CH$_4$, via methanogenesis, in K2-18~b was suggested in previous studies \citep{2023FaDi..245...80M, 2023ApJ...956L..13M}. It was also shown previously that an uninhabited Hycean world with a 1 bar surface pressure, as considered by W24, results in significantly lower CH$_4$ abundance in the atmosphere compared to one with a deeper atmosphere, e.g. 100 bar \citep{2023FaDi..245...80M}. The allowed surface pressure for the uninhabited Hycean scenario depends strongly on the dayside albedo which is presently unconstrained. However, if the uninhabited Hycean scenario is indeed incompatible with the CH$_4$ abundance and if, as noted by \cite{2024ApJ...964L..19G}, the W24 mini-Neptune scenario is incompatible with multiple abundances, then the inhabited Hycean scenario emerges as the most viable explanation of the three, with significant implications for habitability. Therefore, it is important to conduct a detailed exploration of the three scenarios to investigate if indeed the retrieved abundances of K2-18~b can be better explained by the mini-Neptune scenario of W24 than the Hycean scenarios.

In what follows, we discuss our methods in Section~\ref{Models}. In Section~\ref{K2-18 b planetary scenarios section} we investigate the mini-Neptune vs Hycean world scenarios for K2-18~b using the photochemical modeling framework of W24 and assess their results. We then conduct a more general exploration of the model parameter space using two photochemical models and our results are presented in Section~\ref{Results section}. We summarise and discuss our conclusions in Section~\ref{Summary and discussion section}.
\renewcommand{\thetable}{\arabic{table}}
\setcounter{table}{0}

\begin{table}[t!]
\centering
\label{Planetary properties table} 
\begin{tabular}{@{}cc@{}}
\toprule
Parameter  & K2-18~b  \\ \midrule
Planetary mass [$M_\oplus$] & 8.63±1.35 [1] \\
Planetary radius [$R_\oplus$] & 2.610±0.087 [1] \\
Bulk density [g cm\textsuperscript{-3}] & $2.67^{+0.52}_{-0.47}$ [1] \\
Instellation [$S_0$] & $1.005^{+0.084}_{-0.079}$ [1]\\
Orbital period [d] & 32.940045±0.000010 [1] \\
 $\textrm{log}_{10}$\ce{(CH4)} mixing ratio & $-1.74^{+0.59}_{-0.69}$ [2] \\
 $\textrm{log}_{10}$\ce{(CO2)} mixing ratio & $-2.09^{+0.51}_{-0.94}$ [2] \\
 $\textrm{log}_{10}$\ce{(H2O)} mixing ratio & $<-3.06$ [2] \\
 $\textrm{log}_{10}$\ce{(NH3)} mixing ratio & $<-4.51$ [2] \\ 
 $\textrm{log}_{10}$\ce{(CO)} mixing ratio & $<-3.50$ [2] \\ 
 \bottomrule
\end{tabular}
\caption{Bulk properties and atmospheric abundances of K2-18~b. For the abundances, reported by \cite{2023ApJ...956L..13M},  we use the retrieved constraints based on their canonical ``One offset" case. References. (1) \cite{2019ApJ...887L..14B}, (2) \cite{2023ApJ...956L..13M}}

\end{table}

\section{Methods}
\label{Models}

In this section, we outline our modeling setup for further exploring photochemistry across the three atmospheric scenarios, using K2-18~b as a case study. We discuss the rationale for the simulations we perform, describe the different initial conditions used, and expand the simulated parameter space in order to examine the influence of additional variables on the predicted atmospheric composition for K2-18~b. 

W24 performed their mini-Neptune atmospheric simulations using a combination of \textsf{PICASO} \citep{2019ApJ...878...70B, 2022ApJ...930...93R, 2023ApJ...942...71M} and \textsf{Photochem} \citep{2023PSJ.....4..169W, 2024ApJ...963L...7W}. \textsf{PICASO} is used to establish the $P$-$T$ profile and initial states in chemical equilibrium in the deeper mini-Neptune atmosphere. \textsf{Photochem} is used for chemical kinetics and photochemistry simulations. For the two Hycean scenarios, W24 used only \textsf{Photochem}. We first use the same setup as W24 for all three scenarios in order to  reproduce their results as described in Section \ref{K2-18 b planetary scenarios section} and shown in Fig.~\ref{Original Wogan cases figure}. In the same figure, we also reproduce the W24 results using \textsf{VULCAN} \citep{2017ApJS..228...20T, 2021ApJ...923..264T} in combination with the chemical equilibrium code \textsf{FastChem} \citep{2018MNRAS.479..865S, 2022MNRAS.517.4070S, 2024MNRAS.527.7263K} for the mini-Neptune case, and using only \textsf{VULCAN} for the Hycean cases. We perform sensitivity tests for each scenario in Section~\ref{Sensitivity to cross sections results} with their respective W24 setup. Additional model runs with \textsf{VULCAN} are carried out and the results are described in Section~\ref{Revisiting K2-18 b scenarios section}. Before detailing the parameters we varied for the simulations, we briefly describe these models.

\subsection{\textsf{Photochem} model}
\label{Photochem model section}

The \textsf{Photochem}\footnote{\href{https://github.com/Nicholaswogan/photochem/tree/main}{https://github.com/Nicholaswogan/photochem/tree/main}} model hails from a previous photochemical model developed over several decades \citep[e.g,][]{1985JGR....9010497K, segura2003ozone}, culminating in the \textsf{Atmos}\footnote{\href{https://github.com/VirtualPlanetaryLaboratory/atmos}{https://github.com/VirtualPlanetaryLaboratory/atmos}} photochemical model \citep{2017ApJ...836...49A} before being adapted into \textsf{PhotochemPy} \citep{2022PNAS..11905618W} and finally \textsf{Photochem}. The \textsf{Photochem} model is described in Appendix C of \cite{2023PSJ.....4..169W}, where it was validated by comparing model outputs with observations from Earth and Titan \citep{2023PSJ.....4..169W}.

W24 implemented some updates to the thermodynamic data and reaction network that was published in \cite{2023PSJ.....4..169W}. In particular, they describe the importance of the \ce{O + H + H} photolysis channel which has a given branching ratio of 0.12 at 121.0 nm (near Lyman-$\alpha$). Inclusion of this channel changed the mixing ratio of \ce{CH4} in the \cite{2021ApJ...921L...8H} 400 ppm--\ce{CO2} model run from $10^{-2}$ to $3 \times 10^{-5}$. This was evidently an important change, but we suggest more updates to the input data are required: Fig.~\ref{Branching ratio figure} shows that large discrepancies exist in branching ratios between different open-source models. Some of the \textsf{Photochem} branching ratios (e.g. for \ce{CH4} and \ce{NH3}) may not be accurate because they are constant with wavelength.

Only \textsf{Photochem} was used for the Hycean scenarios in W24. A climate code within the \textsf{Photochem} software package (which utilises correlated-k radiative transfer) was used to calculate the $P$-$T$ profile for the Hycean world, assuming 30\% of radiation reflected at the top of the atmosphere. What resulted was a $P$-$T$ profile starting with a 320 K surface temperature, which then traced the moist pseudo adiabat in the lower atmosphere at pressures greater than 20 mbar, before reaching a 215 K isothermal stratosphere.

\subsection{\textsf{PICASO} model}
\label{PICASO model section}

\textsf{PICASO}\footnote{\href{https://github.com/natashabatalha/picaso}{https://github.com/natashabatalha/picaso}} is a Python-based 1D atmospheric radiative-convective equilibrium model which enables 1D climate modeling of hydrogen-dominated exoplanet atmospheres and associated synthetic spectroscopy. For their mini-Neptune case, W24 first performed an atmosphere simulation with \textsf{PICASO} to generate a pressure-temperature ($P$-$T$) profile and equilibrium mixing ratios of various molecules. They use a \textsf{Photochem} kinetic simulation between 1 -- 500 bar to compute the chemical equilibrium-to-disequilibrium transition at 1 bar, assuming lower boundary conditions based on chemical equilibrium. Then, connecting the $P$-$T$ profile to an isothermal atmosphere, \textsf{Photochem} was used to predict the concentrations of species between 1 bar and $10^{-8}$ bar with photochemistry implemented. In this photochemistry simulation, the lower boundary conditions at 1 bar were fixed for species with mixing ratios greater than $10^{-8}$ (10 ppbv) from the kinetics simulation.

\subsection{\textsf{VULCAN} model}

\textsf{VULCAN}\footnote{\href{https://github.com/exoclime/VULCAN}{https://github.com/exoclime/VULCAN}} is a photochemical kinetics model for simulating the atmospheres of exoplanets \citep{2017ApJS..228...20T, 2021ApJ...923..264T}. \textsf{VULCAN} can utilise the equilibrium chemistry code \textsf{FastChem}\footnote{\href{https://github.com/exoclime/FastChem}{https://github.com/exoclime/FastChem}} \citep{2018MNRAS.479..865S, 2022MNRAS.517.4070S, 2024MNRAS.527.7263K} to establish initial states in chemical equilibrium. One can initialise runs with a constant vertical mixing ratio instead, and specify fixed fluxes or mixing ratios at the lower boundary. It has recently been used to simulate the mixing ratios of sulfur biosignatures, such as dimethyl sulfide (DMS), over a range of different conditions and UV fluxes, before estimating their detectability \citep{2024ApJ...966L..24T}. Unlike \textsf{Atmos} and \textsf{Photochem} which only uses one value for photochemical cross sections, \textsf{VULCAN} has three separate physical categories for cross sections: photoionisation, photodissociation, and photoabsorption.

\subsection{Simulation setup}

We first use the modeling setup of W24 to reproduce their results. We then also reproduce these results using \textsf{VULCAN}, before altering some assumptions in both models to achieve different results to W24. With each model, we perform simulations of all three scenarios: mini-Neptune, uninhabited Hycean, and inhabited Hycean. 

The simulations using the W24 setup are shown in Table~\ref{Boundary condition table W24 setup}. For each scenario with the W24 setup, we vary the assumed photochemical cross sections (see Appendix \ref{Appendix: Cross sections and branching ratios}) and the stellar spectrum (see Appendix \ref{Appendix: The assumed stellar spectrum}). In the mini-Neptune setup, we additionally vary the metallicity ($30 - 200\times$ solar), the C/O ratio ($0.25 - 2\times$ solar), and the internal temperature, $T_\textrm{int}$, of the exoplanet ($30 - 70$ K).

The convergence criterion in \textsf{VULCAN} is defined in section 3 of \cite{2017ApJS..228...20T} by equations 10 and 11. If the variation in the model is less than a given factor over a defined period of model time, then the convergence criterion is satisfied and the model integration stops. However, \textsf{VULCAN} can sometimes appear to converge and conclude the run before steady state has been achieved in the model (Nick Wogan, Shami Tsai; private communication). Where necessary, we overrode this steady state criterion by setting \textit{yconv\textunderscore cri = 0}  and \textit{yconv\textunderscore min = 0} in the file \textit{vulcan\textunderscore cfg.py.}

The \textsf{VULCAN} simulations are listed in Table \ref{Boundary condition table VULCAN}. In the mini-Neptune scenario, we perform simulations over different metallicities and compare to the W24 results. For the uninhabited Hycean scenario, we extend the surface pressure to explore the effect it has on the resultant composition. The mini-Neptune simulation converges relatively quickly compared to the Hycean scenarios. Therefore, in the Hycean cases, we ran \textsf{VULCAN} for a total of $10^{17}$ s ($\sim 3.2$ Gyr), noting here that the estimated age of K2-18 is $2.4 \pm 0.6$ Gyr \citep{2019RNAAS...3..189G}. This simulation time accounts for a conservative upper estimate of the age of the system in order to capture the possible long atmospheric lifetime of some molecules.

Initial conditions, including the abundances of chemical species and elemental ratios, can impact the final calculated concentrations of molecules. For the W24 uninhabited Hycean case, the final \ce{CH4} mixing ratio is $3 \times 10^{-10}$, assuming that the atmosphere starts with essentially no \ce{CH4} and accumulates it through photochemical reduction of \ce{CO2} to \ce{CH4}. However, the \ce{H2}-rich atmosphere of K2-18~b is unlikely to form with negligible amounts of \ce{CH4}, so we test different starting abundances and different initial metallicities in \textsf{VULCAN} for the Hycean scenarios. On Hycean exoplanets, the \ce{CO2} abundance in the atmosphere is set by ocean chemistry \citep{2018ApJ...864...75K, 2021ApJ...921L...8H}, rather than initial chemical equilibrium conditions in the atmosphere. Therefore, in line with previous work \citep{2024ApJ...966L..24T} including W24, for the Hycean cases, we fix the \ce{CO2} abundance to ensure it is consistent with retrieved atmospheric abundances. For the inhabited Hycean scenarios, we vary the \ce{CH4} flux at the lower boundary (1 bar) to represent methanogenesis from methanogens and implement a fixed deposition velocity of $1.2\times10^{-4}$ cm s\textsuperscript{-1} as used in W24 to represent acetogens consuming CO. 

We also perform simulations with \textsf{VULCAN} to ascertain whether any mini-Neptune cases can match the retrieved constraints for K2-18~b. We assume the same properties that W24 did for K2-18~b and assume the same setup with \textsf{FastChem} and \textsf{VULCAN} to recreate the \textsf{PICASO} and \textsf{Photochem} simulation. We use \textsf{FastChem} in \textsf{VULCAN} in order to determine the initial chemical composition of the exoplanet's atmosphere, based on the $P$-$T$ profile and given elemental abundances relative to hydrogen, between 500 bar and 1 bar. Like W24, we perform a kinetics simulation between 500 -- 1 bar until the code converged. A $K_\textrm{zz}$ of $10^{8}$ cm\textsuperscript{-2} s\textsuperscript{-1} was used between 500 -- 1 bar. Then, using photochemistry with no top of atmosphere albedo, we simulate until convergence between 1 bar and $10^{-8}$ bar, with a vertically varying $K_\textrm{zz}$ which starts at $10^{3}$ cm\textsuperscript{-2} s\textsuperscript{-1} at 1 bar as in W24. In the \textsf{VULCAN} simulations, we use the chemical networks labelled ``NCHO\textunderscore photo\textunderscore network.txt'' and ``SNCHO\textunderscore photo\textunderscore network\textunderscore  2024.txt'' for the Hycean and mini-Neptune scenarios, respectively, unless otherwise specified \citep[some sulfur species may be sequestered in the ocean on a Hycean world;][]{2021ApJ...921L...8H, 2019ApJ...887..231L, 2024ApJ...966L..24T}.

Molecular photoabsorption, photodissociation, and photoionisation cross sections which vary with wavelength are key parts of the input data for any photochemical model. Photodissociation cross sections quantify the likelihood of a photon initiating a photochemical reaction in a particular chemical species. Accurately representing cross sections is essential for predictions regarding exoplanetary atmospheres and the transmission, reflection, or emission spectra that could be observed from afar. For 22 molecules, including \ce{H2}, \ce{H2O}, \ce{CH4}, \ce{CO2}, \ce{CO}, and \ce{NH3}, the input data is different between three open-source codes: \textsf{Photochem}, \textsf{Atmos}, and \textsf{VULCAN} (see Appendix \ref{Appendix: Cross sections and branching ratios} for more details). Either one set is accurate, or all are inaccurate, and many of the cross section sources in \textsf{Photochem} and \textsf{Atmos} are unknown or not listed. For all three scenarios with the W24 setup, we swap in the cross section data from \textsf{VULCAN} and \textsf{Atmos} with the \textsf{Photochem} data to determine the effect on predicted final composition.

\begin{figure*}[t!]
	\centering
	\includegraphics[width=1\textwidth]{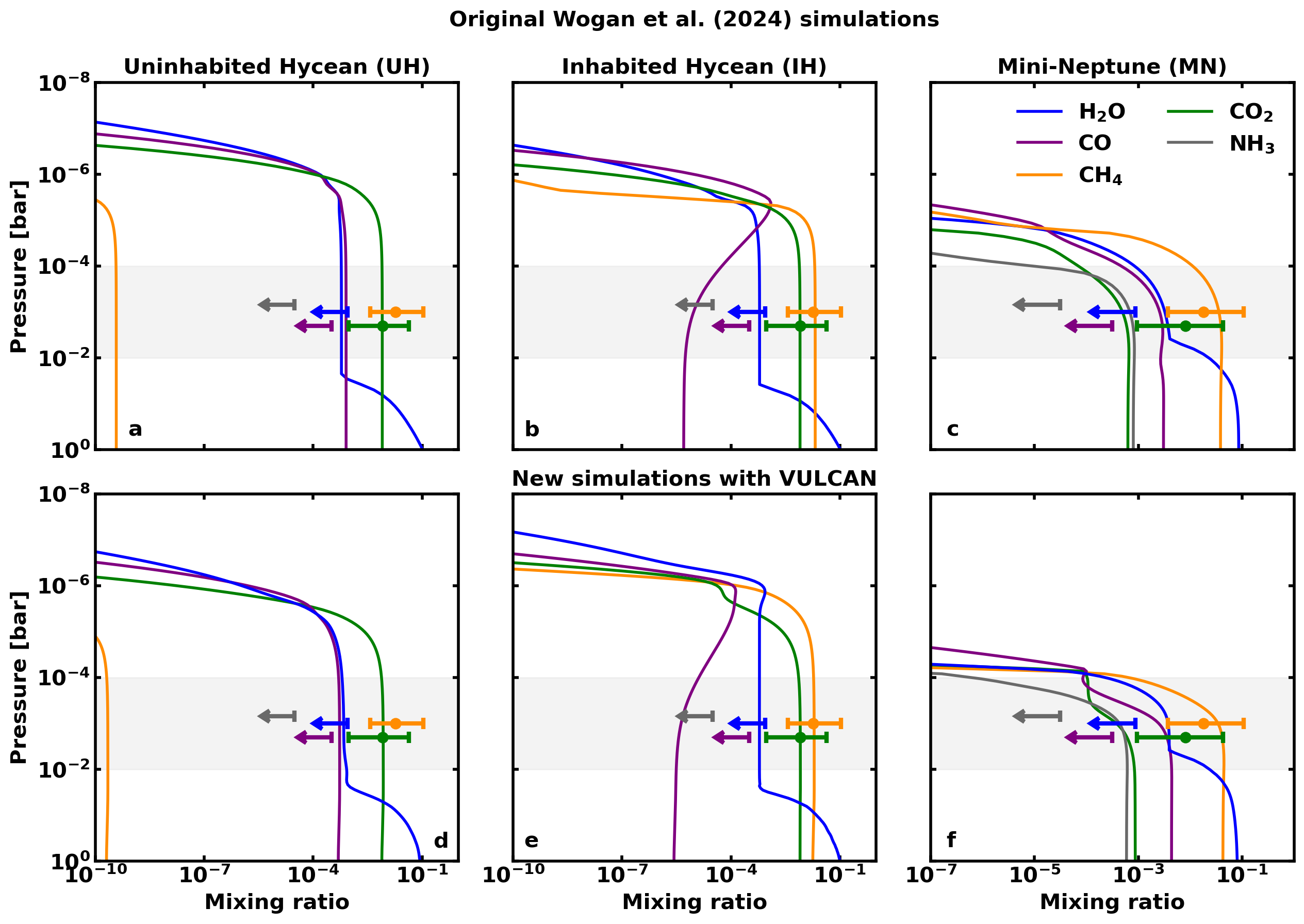}
    \caption{Hycean and mini-Neptune scenarios with two photochemical models. Top row: Original simulations from \cite{2024ApJ...963L...7W}, showing the mixing ratio of molecules against atmospheric pressure in bar. Bottom row: \textsf{VULCAN} simulations of the same scenarios with equivalent assumptions (see main text). The scenarios are: an uninhabited Hycean (left column), an inhabited Hycean (middle column), and a mini-Neptune case (right column). The given molecular constraints are from the ``One offset" case \citep{2023ApJ...956L..13M} as this case gives the best fit to the observed data. 1-$\sigma$ error bars are given for detected molecules (\ce{CH4} - orange; \ce{CO2} - green), and 2-$\sigma$ upper limits are given for \ce{CO} (purple), \ce{H2O} (blue), and \ce{NH3} (grey), which were not detected by \cite{2023ApJ...956L..13M}. The photosphere, the region in the atmosphere which impacts the transmission spectrum observed, is given in a light grey shade between $10^{-2}$ -- $10^{-4}$ bar. A variety of simulations were run with both the W24 setup (see Table~\ref{Boundary condition table W24 setup}) and \textsf{VULCAN} (see Table~\ref{Boundary condition table VULCAN}). All of the simulations shown in this figure use the GJ~176 spectrum. Only the inhabited Hycean scenarios in both models are consistent with the retrieved abundances.} 
    \label{Original Wogan cases figure}
\end{figure*}

The host star, K2-18, does not have its UV stellar spectra measured precisely. Therefore, several previous studies use the MUSCLES\footnote{\href{https://archive.stsci.edu/prepds/muscles/}{https://archive.stsci.edu/prepds/muscles/}} database \citep[for MUSCLES and Mega-MUSCLES papers, see e.g.,][]{2016ApJ...820...89F, 2016ApJ...824..101Y, 2016ApJ...824..102L, 2019ApJ...871L..26F, 2021ApJ...911...18W}. When it comes to interpreting the retrieved abundances from spectra of JWST measurements, we aim to demonstrate why this method introduces additional uncertainties (see appendix \ref{Appendix: The assumed stellar spectrum} for more details) by implementing three stellar proxies for K2-18 in all of the photochemical simulations.

\section{K2-18 b: Hycean or Mini-Neptune?}
\label{K2-18 b planetary scenarios section}

As discussed above, W24 investigate the atmospheric chemistry of K2-18~b under different planetary scenarios, mini-Neptune vs Hycean. The mini-Neptune case corresponds to a deep (500 bar) H$_2$-rich atmosphere whereas the Hycean cases consider a shallow 1-bar atmosphere overlying an ocean, leading to different atmospheric/surface conditions. In what follows, we discuss the model assumptions and conclusions of W24, reproduce their results with two different modeling frameworks, and comment on the Hycean vs mini-Neptune scenarios from their work.

\subsection{Comparison with Retrieved Abundances}

W24 conclude that their predicted abundances for the mini-Neptune case are broadly consistent with the retrieved abundances for K2-18~b. Therefore, we first assess how the chemical abundances predicted by the photochemical models of W24 for the different scenarios compare with the retrieved abundances for K2-18~b \citep{2023ApJ...956L..13M}. As previously mentioned, \cite{2024ApJ...964L..19G} noted that the \ce{NH3} and \ce{CO} mixing ratios predicted by W24 for the mini-Neptune scenario are too high compared to the retrieved abundances. We show the comparison between the predicted abundances for all three W24 planetary scenarios and the retrieved values in the top row of Fig.~\ref{Original Wogan cases figure}. Our reproduction of the W24 scenarios using a separate 1D photochemical code, \textsf{VULCAN}, is shown in the bottom row of Fig.~\ref{Original Wogan cases figure}. 

Using the assumptions of W24, the only planetary scenario that matches all five of the retrieved chemical abundances from the best-fit case in \cite{2023ApJ...956L..13M} in the observable photosphere is the simulated inhabited Hycean scenario. The uninhabited Hycean scenario matches the constraints for \ce{H2O}, \ce{NH3}, and \ce{CO2}, with \ce{CO} on the borderline just above the upper limit, and \ce{CH4} several orders of magnitude too low. In the mini-Neptune scenario, only \ce{CH4} actually matches the 1-$\sigma$ constraints, with \ce{CO2} just below the 1-$\sigma$ constraint, whilst \ce{H2O}, \ce{NH3}, and \ce{CO} have greater abundances than the 2-$\sigma$ upper limits. Therefore, contrary to the claim in W24, their mini-Neptune scenario is not consistent with the retrieved abundances.

\subsection{Initial conditions}

We now assess the initial conditions used in W24 between the three scenarios and their influence on the predicted abundances. In their Hycean simulations, W24 assume that \ce{CH4} starts at negligible abundances, and increases primarily through a biogenic methane flux from the ocean in the inhabited case or through photochemical reduction of \ce{CO2} to \ce{CH4} in the uninhabited case. For the inhabited Hycean case, the biogenic flux is a free parameter which can be tuned to match the retrieved value; therefore, its agreement with the retrievals is expected by default. The uninhabited Hycean case was disfavoured by W24 because insufficient \ce{CH4} was produced photochemically, and the initial CH$_4$ abundance was assumed to be negligible, which is unlikely to be the case in a temperate \ce{H2}-rich atmosphere. On the other hand, in the mini-Neptune case in W24, the initial \ce{CH4} abundance was set by the assumed $100\times$ solar metallicity, which naturally results in a high abundance that matches the retrieved value. Finally, W24 did not test higher surface pressures than 1 bar for the Hycean world scenarios, while higher surface pressures are expected to result in a greater \ce{CH4} abundance \citep{2021ApJ...914...38Y, 2023FaDi..245...80M}. 

\subsection{Temperature structure and albedo} 

\begin{figure}[b!]
	\centering
 \includegraphics[width=\columnwidth]{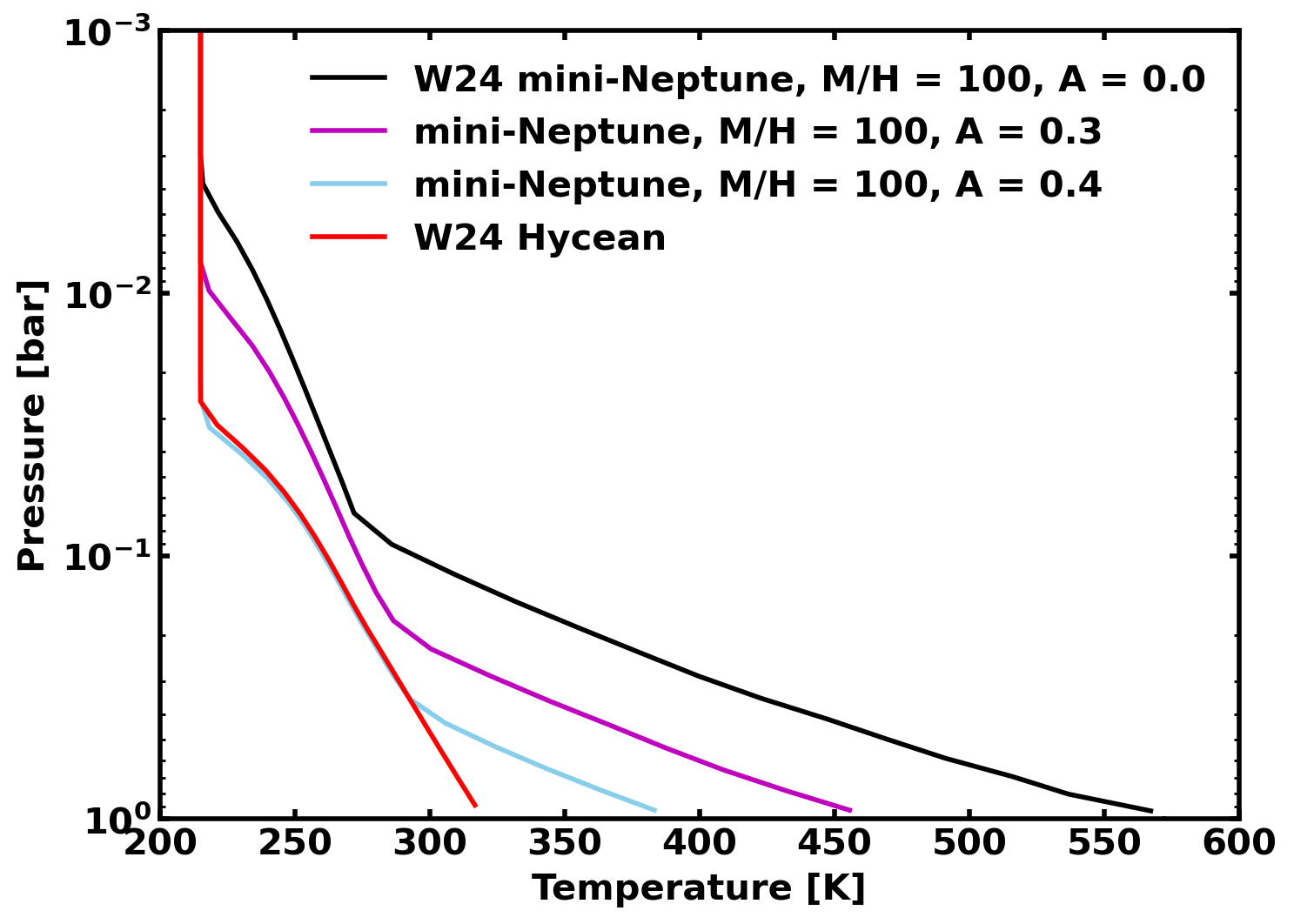}
    \caption{$P$-$T$ profiles for Hycean and mini-Neptune scenarios. The two pressure-temperature profiles that are used in W24, for the Hycean (red) and mini-Neptune (black) atmospheric simulations at pressures less than 1 bar. At pressures less than $3\times10^{-3}$ bar, the temperature is at 215 K in both models. A second and third mini-Neptune model that we simulated shows the mini-Neptune temperature profile when 30\% and 40\% of radiation is reflected off the top of the atmosphere, shown in magenta and light blue, respectively.}
    \label{Wogan temperature structure figure}
\end{figure}

\begin{figure}[t!]
	\centering \includegraphics[width=1\columnwidth]{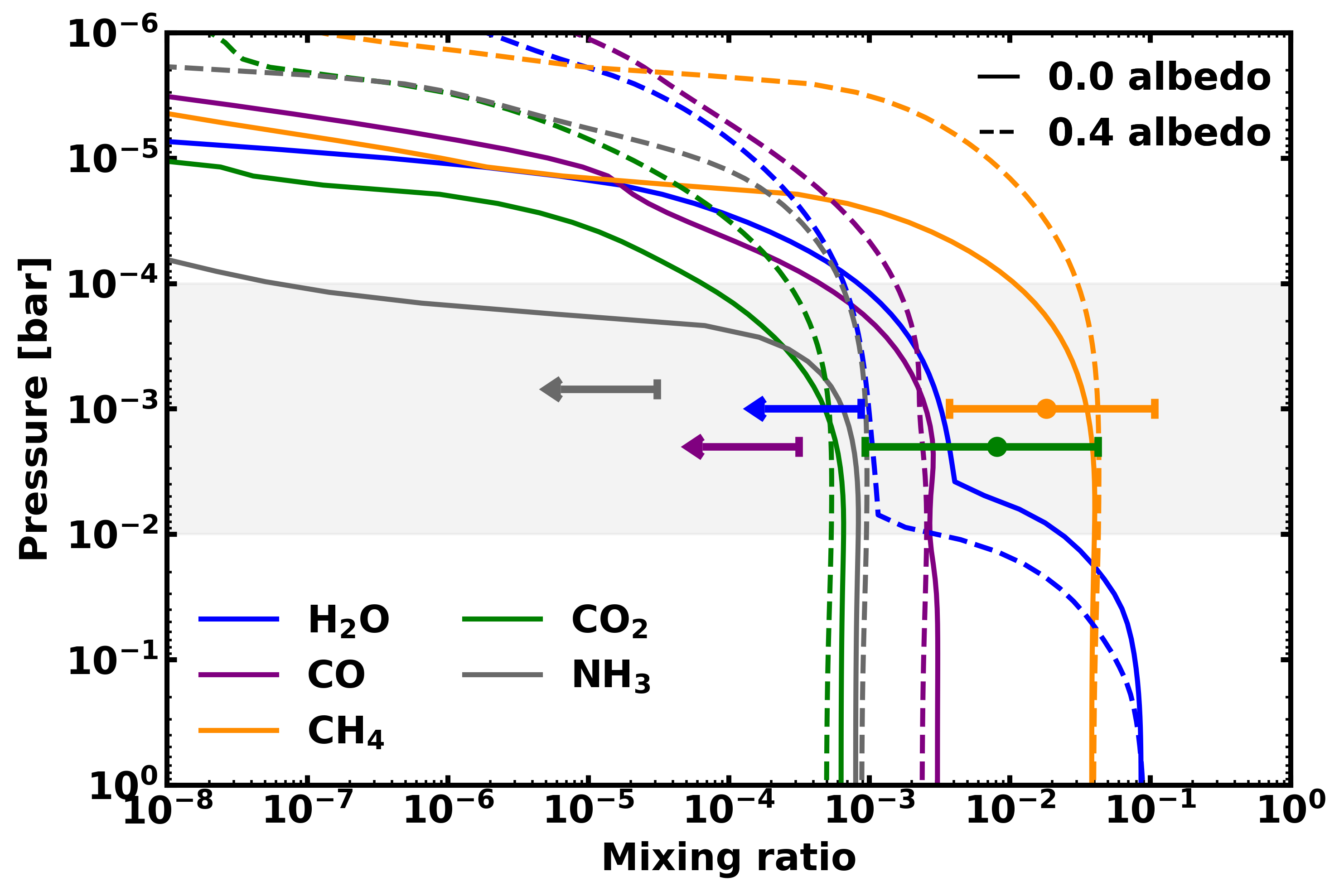}
    \caption{Mini-Neptune scenario with different Bond albedos. These simulations used the W24 setup for the mini-Neptune scenario for K2-18~b, showing the standard $100\times$ solar metallicity scenario with no albedo by the solid lines and the 0.4 top of atmosphere albedo case in the dashed lines. The photosphere is marked in light grey. 1-$\sigma$ constraints are given for detected molecules (\ce{CH4} and \ce{CO2}) using the one-offset values from \cite{2023ApJ...956L..13M}, whilst 2-$\sigma$ upper limits are given by the arrows for \ce{NH3}, \ce{H2O}, and \ce{CO}. In the 0.4 albedo case, \ce{H2O} is on the border of the upper limit, with \ce{CO2} is still lower than its 1-$\sigma$ limit, and CO and \ce{NH3} remain too high.}
    \label{Mini-Neptune albedo figure}
\end{figure}

The predicted abundances in photochemical models are naturally sensitive to the assumed temperature structure in the atmosphere. Fig.~\ref{Wogan temperature structure figure} shows the different $P$-$T$ profiles used for the mini-Neptune and Hycean cases from W24. At pressures between $3\times10^{-3}-1$ bar, the two profiles significantly diverge. As can be seen in other works, which either move the surface pressure between simulated scenarios or assume a deep atmosphere with no surface \citep[e.g.,][]{2021ApJ...922L..27T, 2021ApJ...914...38Y}, the $P$-$T$ profiles should be consistent when they are at pressure levels shared between the simulations for a given albedo. This is because the presence of a surface does not significantly alter the $P$-$T$ profile in the upper atmosphere \citep{2021ApJ...922L..27T}.

The Bond albedo plays a central role in driving both the radiation field in a photochemical model as well as the temperature profile. The observed transmission spectrum of K2-18~b provides evidence for the presence of clouds/hazes in the planet \citep{2023ApJ...956L..13M}. W24 assume different prescriptions for the incoming radiation between the Hycean and mini-Neptune cases. W24 reduce the incoming radiation by 30\% in the Hycean case and not at all in the mini-Neptune case. We also include two additional mini-Neptune temperature profiles in Fig.~\ref{Wogan temperature structure figure} when assuming that 30\% and 40\% of radiation is immediately reflected off the top of the atmosphere. A 40\% albedo is required to match the temperature profile in the W24 Hycean case at pressures less than 0.4 bar. This albedo of 0.4 is also required for \ce{H2O} to just be consistent with the retrieved upper limit for \ce{H2O}, with \ce{CH4} being the only other molecule that matches the molecular constraints.

The 30\% of the incoming radiation would not be reflected at the top of the atmosphere in reality but instead reduced through scattering, photoabsorption and photodissoication, when passing through denser atmospheric layers. Moreover, W24 use the GJ~176 spectrum to act as a proxy for K2-18 \citep[following][]{2020ApJ...898...44S, 2021ApJ...921L...8H}. GJ~436 has been used in other work \citep{2024ApJ...966L..24T}, and in Appendix \ref{Appendix: The assumed stellar spectrum} we suggest that GJ~849 could also be used. So, different simulations in the literature use different spectra and alter the incoming radiation through different methods. All these factors influence the propagation of ultraviolet (UV) radiation and affect the resulting atmospheric chemical composition differently between the Hycean and mini-Neptune cases.

\subsection{Synthetic spectra} 
\label{Synthetic spectra}

W24 stated that both a mini-Neptune case and an inhabited Hycean case can explain the observations from \cite{2023ApJ...956L..13M}. To reach this conclusion, they generated transmission spectra of forward models from the photochemical results and calculated a reduced $\chi^2$ ($\chi^2_r$) for each scenario. The $\chi^2_r$ test assumed that the number of degrees of freedom was the number of data points (i.e. no free parameters). For the uninhabited Hycean, the inhabited Hycean, and the mini-Neptune clear-sky scenarios, a $\chi^2_r$ of 3.22, 1.51, and 1.51, respectively, was found. When aerosols were included in the scenarios, the $\chi^2_r$ for each scenario was 2.30, 1.40, and 1.46, respectively. Then, after removing the JWST data shortward of 1 $\mu$m, the $\chi^2_r$ was 2.10, 1.15 and 1.15, respectively. 

This approach is reminiscent of the pre-retrieval era whereby individual forward models were fit to low-resolution exoplanet spectra which precluded a robust assessment of model degeneracies and confidence estimates \citep{2009ApJ...707...24M}. For example, it is unclear how many other photochemical models would match at that $\chi^2_r$ reported by W24 or at a better $\chi^2_r$, such that there is no evaluation of whether the forward models are unique solutions. Furthermore, not accounting for the various free parameters in each model scenario biases the model comparisons. The $\chi^2_r$ as presented by W24 could be seen as a measure of the goodness-of-fit of the specific parameter choices within each model scenario. However, it does not serve as evidence to assess whether one model scenario explains the data better than another. For example, a different instance in the uninhabited Hycean model space may provide a better $\chi^2_r$ than reported, but even then it would not provide sufficient grounds for comparison between that model scenario and another. A reliable approach to evaluate model scenarios is to compare either (a) the best-fit $\chi^2_r$ over the full parameter space of each model scenario, accounting for the true number of degrees of freedom, or (b) more accurately, the Bayesian evidence by integrating the prior-weighted likelihood over the model parameter space, as computed in extant atmospheric retrievals \cite[e.g.,][]{2023ApJ...956L..13M}.

The procedure does also not reveal why the different synthetic spectra achieved the same $\chi^2_r$, despite having significantly different abundances of molecules in their simulated atmospheres. It is also possible to fit an $n^{th}$ degree polynomial to the data which achieves a similar $\chi^2_r$, but this reveals nothing about the atmospheric molecular abundances because such a fit is not based on a physically plausible model, unlike atmospheric retrievals. This is why atmospheric retrievals are implemented to provide confidence estimates on specific variables based on a free parameter space exploration of millions of individual models. Therefore, photochemical model results should instead be compared to the retrieved abundances, especially in the JWST era where high-quality spectra allow for detailed Bayesian inferences.

\subsection{Occam's Razor for Model Preference}

The primary conclusions of W24 are that (a) an uninhabited Hycean scenario in K2-18~b is inconsistent with the observations, (b) the inhabited Hycean and mini-Neptune scenarios are in comparable agreement with observations, and (c) the mini-Neptune scenario is preferred over the inhabited Hycean through an Occam's Razor argument that the mini-Neptune model is simpler. However, based on the above discussion, the mini-Neptune scenario is in least agreement with the retrieved abundances, while the Hycean scenarios perform better, rendering Occam's Razor inapplicable in the present context. Nonetheless, we evaluate the arguments of W24 in favour of the mini-Neptune scenario and against the Hycean scenario.

W24 state that a mini-Neptune with a 60 K intrinsic temperature, and an atmosphere that has 100$\times$ solar metallicity and a solar C/O ratio, can broadly explain the observed \ce{CH4} and \ce{CO2} abundances. However, their calculated \ce{CO2} actually lies at a lower abundance than the 1-$\sigma$ retrieved limit. An intrinsic temperature of 60 K was chosen based on the modelling from \cite{2021ApJ...921...27H}, who stated that this value is similar to the internal temperature of Neptune. The internal heat of Neptune was estimated to be $0.433 \pm 0.046$ W m\textsuperscript{-2} \citep{1991JGR....9618921P}, which corresponds to an internal temperature of $52.57$ K. This internal temperature estimate was derived from Voyager observations of Neptune, which may need to be updated in light of the Cassini results for Jupiter \citep{li2018less}. Additionally, the intrinsic temperature of a mini-Neptune is unknown and the internal temperature of K2-18~b could be lower than 60 K \citep{2013ApJ...775...10V, 2020ApJ...891L...7M}: internal temperature generally scales with planet mass \citep{2012A&A...547A.111M, 2014ApJ...792....1L}; Neptune is roughly twice the mass of K2-18~b; and GJ 1214 b, which has a comparable mass to K2-18~b (within 10\%), has been suggested to have an internal temperature of 30 K \citep{2013ApJ...775...10V}. Given a colder internal temperature, then \ce{CO2} would be even less abundant\footnote{We tested the effect of internal temperature with the standard 100 times solar metallicity model from W24. For internal temperatures of 52.6 K, 50 K, 40 K, and 30 K, the \ce{CO2} mixing ratio at 1 mbar was $3\times10^{-4}$, $2\times10^{-4}$, $5\times10^{-5}$, and $2\times10^{-5}$, respectively. The 1-$\sigma$ lower limit on \ce{CO2} in the `one offset' case is $9\times10^{-4}$}.

W24 claim that the simulated deep-atmosphere kinetics produces abundances of \ce{CO} and \ce{NH3} which are generally compatible with their non-detection. In fact, at 1 mbar, \ce{NH3} and \ce{CO} are both calculated to be at greater abundances (by a factor of 23, and 8 respectively) than their retrieved 2-$\sigma$ upper limits.

The absence of \ce{H2O} features in the transmission spectrum are attributed to water vapor condensation and cold trapping by W24. We agree with this statement, but this is not what the simulated mini-Neptune scenario from W24 shows: the predicted \ce{H2O} mixing ratio is a factor of 4 too high at 1 mbar, and a factor of 28 too high at 10 mbar, compared to the 2$-\sigma$ upper limit on \ce{H2O}. In the mini-Neptune scenario using the W24 setup, to get enough condensation to explain the non-detection of \ce{H2O}, we find that a top of atmosphere albedo of $\geq 0.4$ is required by the W24 setup, although CO and \ce{NH3} remain too high and \ce{CO2} remains too low in abundance (see Fig.~\ref{Mini-Neptune albedo figure}).

W24 also claim that 1D radiative-convective-equilibrium modeling can explain the climate of K2-18~b in the case of the mini-Neptune. This is despite no observational constraints on the temperature structure of the dayside atmosphere or on radiative-convective equilibrium for K2-18~b. As a climate argument against the Hycean scenario, it has been discussed that a Hycean exoplanet could undergo a steam runaway greenhouse \citep{2020ApJ...898...44S, 2023ApJ...953..168I, 2023ApJ...944...20P} unless the Bond albedo is sufficiently high in order to reflect enough incoming radiation \citep{2020ApJ...904..154P, 2021ApJ...918....1M}. For K2-18~b to be a Hycean world, most recent dayside 3D simulations require a Bond albedo of $\sim 0.5 - 0.6$ \citep{2024A&A...686A.131L}. Such an albedo cannot be ruled out given the inference of clouds/hazes in K2-18~b \citep{2023ApJ...956L..13M}, and several planets, including Jupiter, have Bond albedos reported between $\sim 0.5$ -- $0.7$ \citep{li2018less, 2020ApJ...903L...7C, 2023Natur.620...67K, 2023A&A...675A..81H}. 

While a sufficiently high Bond albedo to maintain an ocean on K2-18~b is not implausible, a robust albedo estimate is currently not available. \cite{2024A&A...686A.131L} (L24) use model spectra to assess the dayside albedo of K2-18~b based on the two weakest CH$_4$ features in the transmission spectrum (at 1 $\mu$m and 1.2 $\mu$m). The observed amplitude of one such feature is used to infer its haze properties at the day-night terminator and then extrapolate the findings to the dayside albedo. At the outset, it is unclear if any of the L24 models are compatible with the observed transmission spectrum, as no such comparison is presented. Secondly, heuristic metrics based on such limited weak spectral features and nominal model considerations (e.g. homogeneous Rayleigh-like hazes) are unlikely to obtain robust constraints on clouds/hazes and resolve various model degeneracies, e.g. between clouds/hazes, molecular contributions and temperature. This is better pursued with atmospheric retrievals with the full data available (1-5$\mu$m), which indeed show evidence for clouds/hazes as discussed above. It is also important to recognize that the scattering on the dayside can be very different to that at the terminator, rendering extrapolations to dayside albedo indicative at best. Finally, inferences of clouds/hazes and Bond albedos have to be conducted within a self-consistent framework that also considers all available chemical abundance constraints. The models of K2-18~b reported by L24 predict \ce{NH3} and CO abundances 1-2 orders of magnitude higher than the retrieved $2$-$\sigma$ upper limits. Future observations and self-consistent cloud/haze modeling would be helpful to obtain better insights into the possible dayside Bond albedo of K2-18~b.

Finally, W24 argue that XUV-driven hydrogen escape may erode a thin $\sim1$ bar \ce{H2}-dominated atmosphere that cannot be replaced by volcanism \citep{2016Icar..277..215N, 2018ApJ...864...75K, 2023ApJ...948L..20H}. This is indeed a possibility, however, there is no quantitative analysis that supports such a conclusion for K2-18~b specifically, considering a wide range of initial and steady state conditions that may be possible on K2-18~b. It is feasible that the primordial atmosphere may have been significantly deeper than the present state. Furthermore, given an adequate albedo, K2-18~b could have a deeper atmosphere (e.g., 10 bar) and still maintain Hycean conditions \citep{2020ApJ...904..154P, 2023FaDi..245...80M}. For \ce{CH4} to be present in the observed abundance, W24 argue that K2-18~b as a Hycean world requires a biogenic, or other, source of \ce{CH4}. This assumes that the surface pressure of the Hycean world scenario is 1 bar, which is just one possibility given all the current unknown physical parameters regarding K2-18~b.

Overall, we find the conclusions of W24 to be inconsistent with the retrieved abundances and dependent on initial conditions between the different model scenarios. Nevertheless, it is important to further investigate the viability of the Hycean vs mini-Neptune scenarios. We therefore revisit the two Hycean scenarios (inhabited and uninhabited) and the mini-Neptune scenario for K2-18~b through further photochemical modeling with two independent photochemical models and expand the simulated parameter space to determine which scenarios can best explain all the retrieved abundance constraints.

\section{Results}
\label{Results section}

In this section, we reexamine the mini-Neptune scenario and the two Hycean scenarios for K2-18~b using both the W24 photochemical modeling setup and the \textsf{VULCAN} photochemical model. We explore a range of initial conditions for the mini-Neptune scenario to cover a broader parameter space, aiming to identify any simulated cases that can explain the observed abundance constraints. Additionally, we investigate various parameters, such as photochemical cross sections and the stellar input spectrum, to assess the sensitivity in calculated atmospheric composition using the W24 setup for all three scenarios. Finally, we increase the surface pressure and test the results for the uninhabited Hycean scenario. 

\subsection{Sensitivity to Cross Sections and Stellar spectra}
\label{Sensitivity to cross sections results}

For both mini-Neptune and Hycean scenarios with the W24 setup, we explore different input data and initial conditions to determine the effects on the calculated mixing ratio profiles for various molecules of observational and biological interest. We find significant differences in the photochemical cross sections between three publicly available photochemical models: \textsf{Atmos}, \textsf{Photochem}, and \textsf{VULCAN}, as shown in Appendix \ref{Appendix: Cross sections and branching ratios}. 

We recompute the W24 scenarios using the same model setup but replacing the photochemical cross sections with those from different sources. In this section, we present the results from these cross section comparisons. We found that there were discrepancies for twenty-two molecules which we swapped into the W24 setup. Because the wavelength binning of the cross sections are sometimes different between sources and molecules, we vary the wavelength binning of the cross sections, keeping the original resolution (native) as well as binning to the resolution of the cross sections for \textsf{Photochem} molecules (binned). Furthermore, we test the impact of three different stellar spectra: GJ~176, GJ~436, and GJ~849, which have different strengths and shapes of incoming UV radiation (see Appendix \ref{Appendix: The assumed stellar spectrum} for a discussion on stellar proxies and Fig.~\ref{Stellar spectra figure} and Fig.~\ref{Fractional stellar flux figure}).

\subsubsection{Mini-Neptune scenario}

\begin{figure*}[t!]
	\centering \includegraphics[width=0.95\textwidth]{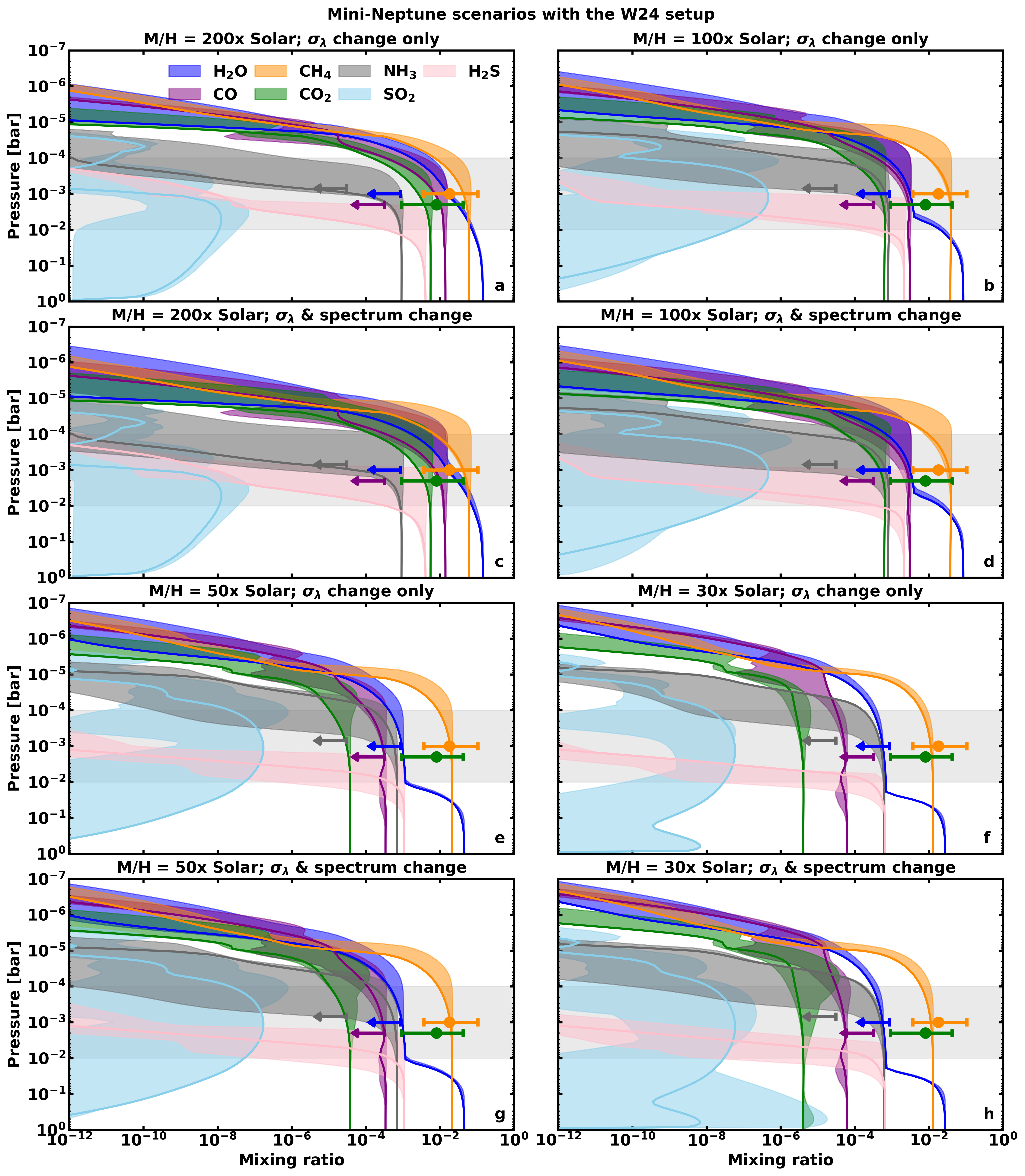}
    \caption{Mini-Neptune scenario for K2-18~b using the W24 setup (\textsf{PICASO} and \textsf{Photochem}). The predicted mixing ratios for different metallicities ($200\times$, $100\times$, $50\times$, $30 \times$ solar metallicity, denoted by $M/H$) are shown by the shaded regions, indicating the maximum deviations between the different simulations. The various simulations used different cross section sources and resolutions in the first and third row, assuming the host star is represented by GJ~176, and the internal temperature ($T_\textrm{int}$) is 60 K. The second and fourth row show the same, but this time including all the same simulations but with GJ~176, GJ~849, and GJ~436 as the host stars. The photosphere is marked in light grey. 1-$\sigma$ constraints are given for detected molecules (\ce{CH4} and \ce{CO2}) using the values from \cite{2023ApJ...956L..13M}, whilst 2-$\sigma$ upper limits are given by the arrows for \ce{NH3}, \ce{H2O}, and \ce{CO}. The colours used for the constraints are the same as those used in Fig.~\ref{Mini-Neptune albedo figure}, with \ce{SO2} and \ce{H2S} also shown in light blue and pink, respectively. The thick lines in each panel show the results when GJ~176 is used with \textsf{Photochem} cross sections.}
    \label{Mini-Neptune XS star figure}
\end{figure*}

The W24 mini-Neptune case only matched one out of the five abundance constraints, so we performed many more simulations over a larger parameter space to check if other simulations are more consistent with the current observations and atmospheric retrieval analysis.

Using the \textsf{Photochem} and \textsf{PICASO} combination as in W24, we tested different initial conditions on a grid of C/O ratio from 0.5 to 2.0 and metallicity of 30 to 200, a $T_\textrm{int}$ of 30 K to 70 K, and with three different stellar spectra. None of the predicted abundances from any of these cases could match the constraints for all five molecules. Fig.~\ref{Mini-Neptune XS star figure} shows a range of results assuming different initial metallicities with the W24 setup ($30\times, 50\times, 100\times$, and $200\times$ solar metallicity), whereby the cross sections have been altered whilst using either the GJ~176,  GJ~849, or GJ~436 spectrum.

We find that the cross section source, the resolution of the cross sections used, and the assumed stellar spectra, all alter the predicted composition, as shown in Fig.~\ref{Mini-Neptune XS star figure}. When keeping the star constant, and changing the cross sections, peak abundance changes (between 10 and 0.1 mbar in pressure) can be up to a factor of $\sim 3\times10^9$. When keeping the cross sections constant but varying between the three stellar input spectra used, peak abundance changes can be up to a factor of $\sim 1\times10^9$. The maximum differences in calculated abundances in the photosphere at the same pressure level are $3\times10^9$ (\ce{NH3}), $2\times10^9$ (\ce{NH3}), $1\times10^9$ (\ce{H2S}), and $4\times10^7$ (\ce{H2S}), in the $200, 100, 50$, and 30 times solar metallicity cases, respectively.

Peak abundances change for \ce{SO2} by up to a factor of $\sim 10^7$ when using different stellar spectra and cross sections. Other mixing ratio changes occur for \ce{H2O}, \ce{CH4}, \ce{CO2}, and \ce{CO}, although these are smaller in magnitude because the W24 setup fixes their relatively high abundance at the 1 bar lower boundary in the photochemical simulation due to mixing in the deeper atmospheric simulation. The production rate effectively adjusts to compensate for any chemical loss alterations induced by the change in cross section data.

At a greater metallicity than $100$ times solar, \ce{CH4}, \ce{CO2} and \ce{CO} increase at the lower boundary (1 bar) in the photochemical simulations, and the inverse occurs at lower metallicities. When changing the C/O ratio to lower values, carbon bearing species such as \ce{CH4}, \ce{CO2} and CO decrease at the lower boundary, but \ce{H2O} increases, and vice versa when increasing the C/O ratio. 

There are several trends with metallicity and the assumed cross sections used. With increasing metallicity, a change in cross section from the original \textsf{Photochem} cross sections has a larger perturbation effect for the mixing ratios of \ce{CO2}, \ce{CO}, \ce{CH4}, \ce{HCN}, and \ce{H2O}. On the other hand, \ce{H}, \ce{NH3}, and \ce{H2S} have larger mixing ratio perturbations with decreasing metallicity. For these cross section changes, \ce{SO2} is an example where the abundance can be increased or decreased depending on the choice of cross sections and what resolution is used.

The $200 \times$ solar metallicity simulations produce results which become more consistent with the retrieved JWST abundance of \ce{CO2}. This can, however, depend on the source of the cross sections as the \textsf{Atmos} native cross sections significantly decrease \ce{CO2} abundance in the photosphere away from the other profiles, and could mean the difference between remaining consistent or inconsistent with retrieved abundances. Ultimately, whilst certain changes can produce chemical abundances that more closely match the retrieved abundance constraints from \cite{2023ApJ...956L..13M}, none of the mini-Neptune simulations we performed can explain the simultaneous non-detection of \ce{H2O}, \ce{CO}, and \ce{NH3}, and the presence of \ce{CO2} and \ce{CH4} in mixing ratios of $\sim 1$\%. 

Overall, we find that various initial conditions and alternative sets of input data can introduce significant uncertainties on the final predicted atmospheric composition. Such results from different models can then lead to different conclusions and future modeling efforts should be aware of these sensitivities. In the specific mini-Neptune scenario for K2-18~b as simulated here, the sensitivity tests still do not support the mini-Neptune scenario when comparing to retrieved abundances. 

\subsubsection{Hycean scenarios}
\label{Hycean scenarios results}

\begin{figure*}[t!]
	\centering
	\includegraphics[width=0.95\textwidth]{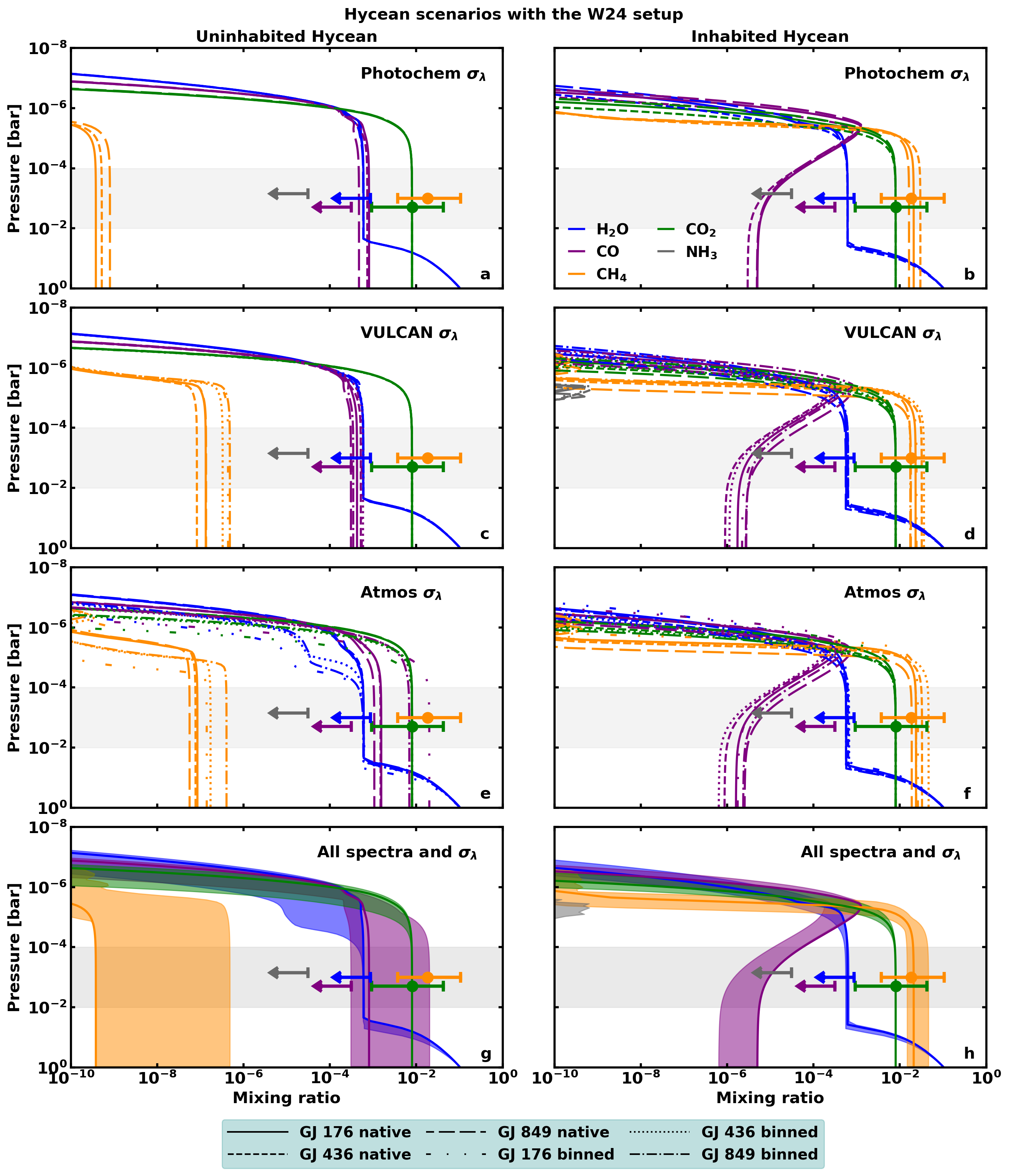}
    \caption{Hycean scenarios for K2-18~b using the W24 setup. The photochemical modelling uses \textsf{Photochem} in the W24 setup for an uninhabited Hycean case (left column) and an inhabited Hycean case (right column). The first, second, and third rows use \textsf{Photochem}, \textsf{VULCAN}, and \textsf{Atmos} cross sections, respectively (see Appendix \ref{Appendix: Cross sections and branching ratios}).  The solid lines, dashed lines and long-dashed lines are with native cross sections using the GJ~176, GJ~436, and the GJ~849 spectra, respectively. The dash-dot-dot-dashed lines, dotted lines, and dash-dotted lines are with binned cross sections using the GJ~176, GJ~436, and the GJ~849 spectra, respectively. The final row shows shaded regions which display the range of results between all perturbations in the top 3 rows. The thick lines in the final row also shows the original W24 results. The ``observable" portion of the atmosphere and the molecular constraints are marked as in Fig.~\ref{Original Wogan cases figure}.}
    \label{Lifeless Inhabited Hycean figure}
\end{figure*}

We now examine the effect of different cross section sources and stellar spectra on the Hycean cases when using the W24 setup. Fig.~\ref{Lifeless Inhabited Hycean figure} shows the uninhabited (left column) and inhabited (right column) Hycean cases with binned and native \textsf{Atmos} and \textsf{VULCAN} cross sections using three different stellar spectra (See Table~\ref{Boundary condition table W24 setup} for a description of simulations performed with the W24 setup). 

In the uninhabited Hycean case with the binned and native \textsf{Atmos} cross sections, \ce{CO} increases at the surface by a factor of 25 and 2 to a mixing ratio of 0.02 and 0.0015, respectively. Yet for the \textsf{VULCAN} cross sections, \ce{CO} decreases by a factor of $1.5$ and $1.9$ when binned or used at their native resolution, respectively. In the inhabited Hycean cases, \ce{CO} decreases by a factor of between 2.1 -- 3.1, depending on the various cross section choices. The \ce{CH4} mixing ratio increases by 360 and 225 for the binned and native \textsf{Atmos} cross sections, and 1150 and 350 times for the binned and native \textsf{VULCAN} cross sections, respectively. 

Generally speaking the W24 inhabited Hycean case appears much less sensitive to the changes in cross sections and stellar spectra when compared to the uninhabited case. This may be because \ce{CH4} photodissociates at UV wavelengths; these photons would otherwise be able to penetrate further into the atmosphere and affect other molecules. When swapping the cross sections in the inhabited hycean scenario, between 1 bar and $10^{-5}$ bar, \ce{CH4} increases by a factor of 1.11 -- 1.25, whereas \ce{CO} decreases by 2.2 -- 3.1 times. \ce{H2O} is essentially unaffected at the surface due to the assumed tropospheric humidity. In the upper atmosphere, \ce{H2O} is significantly perturbed with the binned \textsf{Atmos} cross sections, decreasing by a factor of $\approx 100$ by $10^{-6}$ bar. 

Altogether, we find peak abundances can change in the uninhabited Hycean case for up to $\sim 10^3$ for \ce{CH4} and $\sim 10^2$ for \ce{CO}. In the inhabited Hycean case, \ce{CO} and \ce{CH4} are less affected, altering their abundance by up to a factor of 3. This demonstrates that both the assumed resolution of the cross sections, and the sources the cross sections are from, have a significant impact on the final abundance of specific molecules. The composition of the atmosphere and boundary conditions drive the large difference in predicted discrepancies between the two Hycean scenarios.

Taken in their totality, modeling choices can therefore introduce large uncertainties on the final predicted atmospheric composition if a parameter space is not adequately explored. If we take \ce{CH4} in the uninhabited Hycean case as an example, by only changing the spectrum to GJ~849, one can increase the predicted \ce{CH4} mixing ratio by a factor of 2.1. However, when introducing both a different spectrum (GJ~849) and cross section source (\textsf{VULCAN}), one can increase the \ce{CH4} mixing ratio by a factor of 1260. A modeller will not necessarily know beforehand what the effect of such modifications will be, especially when we have shown that the opposite effect for a specific molecule can occur between the inhabited and uninhabited Hycean scenarios.

\subsection{Revisiting K2-18~b scenarios with \textsf{VULCAN}}
\label{Revisiting K2-18 b scenarios section}

Now we present the atmospheric scenarios using the \textsf{VULCAN} photochemical model, presenting first the mini-Neptune scenario and then the Hycean scenarios, before varying the atmospheric pressure and temperature profile of the uninhabited Hycean scenario in \textsf{VULCAN}. 

\subsubsection{Mini-Neptune scenario}
\label{Mini-Neptune scenario results}

\begin{figure}[t!]
	\centering
 \includegraphics[width=\columnwidth]{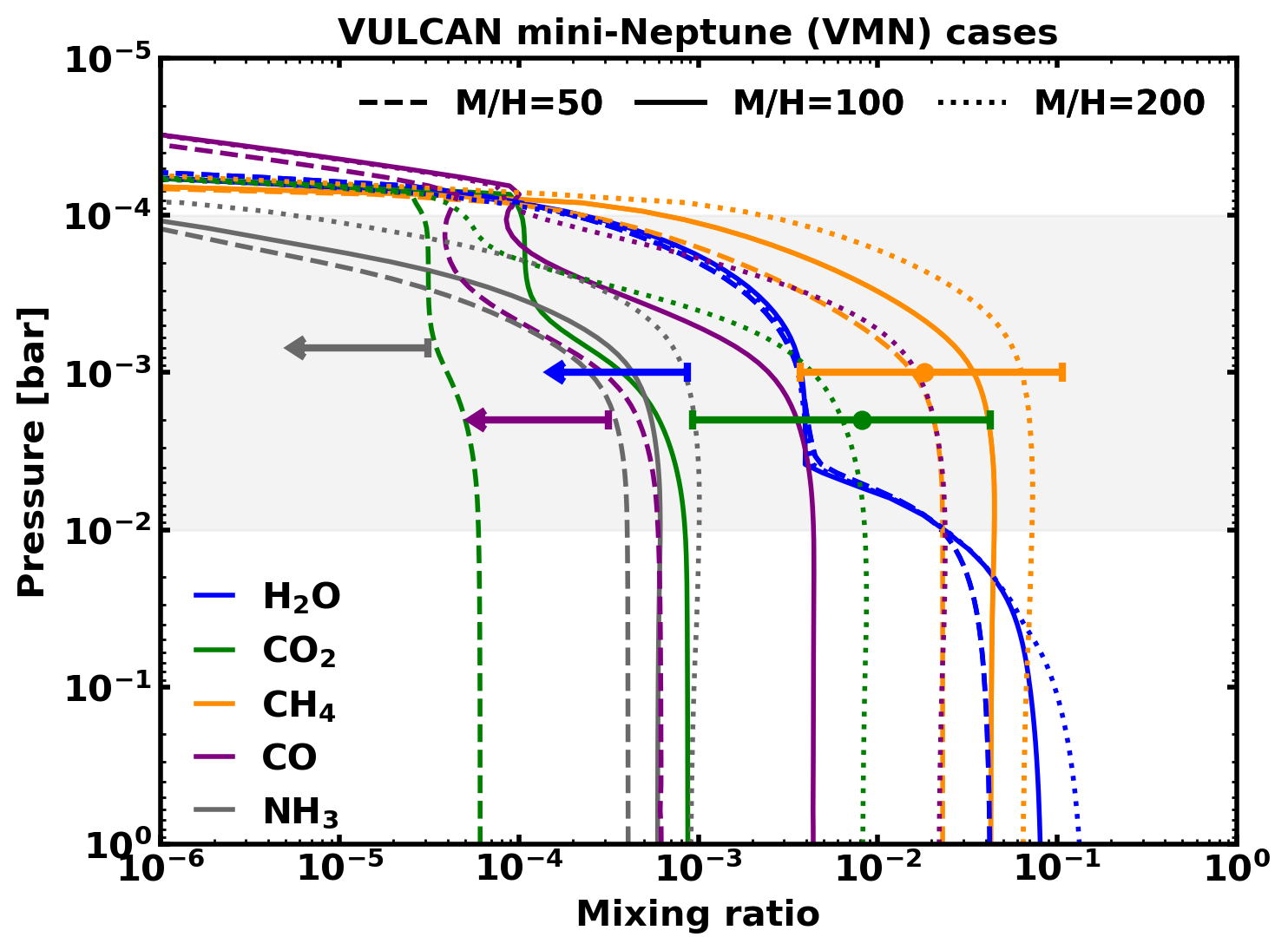}
    \caption{Mini-Neptune scenario for K2-18~b using \textsf{VULCAN}. The mini-Neptune simulations use the same $P$-$T$ and $K_\textrm{zz}$ profiles that were used in W24 from $10^{-8}$ -- 500 bar. The colours and constraints are the same as in Fig.~\ref{Original Wogan cases figure}. The solid line is a $100\times$ solar metallicity case, the dashed line is a $50\times$ solar metallicity case, and the dotted line is a $200\times$ solar metallicity case. None of the cases are consistent with all five retrieved abundances. The predicted abundances are comparable to those produced in using the W24 setup.}
    \label{Mini-Neptune VULCAN figure}
\end{figure}

We set up \textsf{VULCAN} with the W24 $P$-$T$ and $K_\textrm{zz}$ (vertical mixing parameter) profiles and different initial elemental abundances (metallicities), assuming the same properties for K2-18~b. We show examples of $50\times$, $100\times$, and $200\times$ solar metallicity \textsf{VULCAN} mini-Neptune simulations in Fig.~\ref{Mini-Neptune VULCAN figure}. The results are similar in abundance to the W24 results: if \ce{CH4} and \ce{CO2} match the constraints, then there are other species which do not match, including \ce{H2O} and \ce{CO} which have mixing ratios that are greater than the 2-$\sigma$ upper limits. As an illustrative example, \ce{CH4} and \ce{CO2} are consistent with the retrieved abundances in the $200\times$ solar metallicity scenario, but the non-detected molecules are all predicted to be significantly greater than their 2-$\sigma$ upper limits. Additionally, for $50\times$, $100\times$, and $200\times$ solar metallicity, at 1 mbar, \ce{NH3} is a factor of 8, 14, and 27 higher, respectively, than the \ce{NH3} 2-$\sigma$ upper limit.

\begin{figure*}[t!]
	\centering
 \includegraphics[width=1\textwidth]{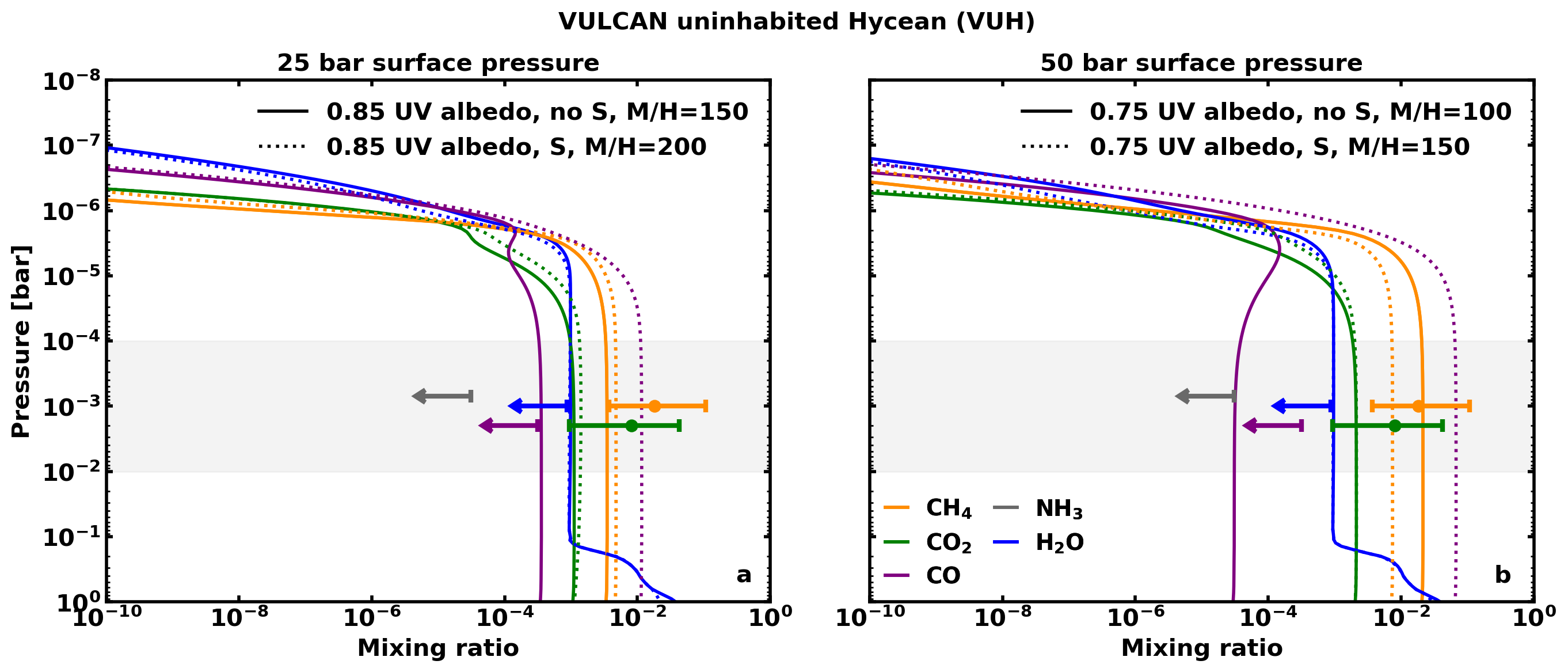}
    \caption{Uninhabited Hycean scenario for K2-18~b using \textsf{VULCAN}. Photochemical modeling of an uninhabited Hycean scenario with a 25 bar surface (a; left) and with a 50 bar surface (b; right). The GJ~436 spectrum with a varied UV albedo is implemented in the model. The $P$-$T$ profile is PT3 from \cite{2023FaDi..245...80M}. Different initial metallicities are used of between 100 -- 200 times solar metallicity. In the uninhabited Hycean scenario, the inclusion of sulfur (labelled S in the legend) in the photochemical network affects the abundance of CO. The uninhabited simulations have no imposed flux for \ce{CH4} at the planetary surface. \ce{CO2} has a fixed mixing ratio condition of $1\times10^{-3}$ (25 bar cases) and $2\times10^{-3}$ (50 bar cases). The mixing ratio of \ce{CH4}, \ce{CO2}, \ce{CO}, \ce{H2O}, and \ce{NH3} are plotted against atmospheric pressure.  The constraints and colours are the same as in Fig.~\ref{Original Wogan cases figure}.  Depending on the assumed physical conditions and the chemical network used, the uninhabited Hycean scenario can be shown to be consistent with the retrieved abundance constraints from JWST observations \citep{2023ApJ...956L..13M} for all five molecules.}
    \label{VULCAN uninhabited Hycean figure}
\end{figure*}

\begin{figure}[b!]
	\centering
 \includegraphics[width=\columnwidth]{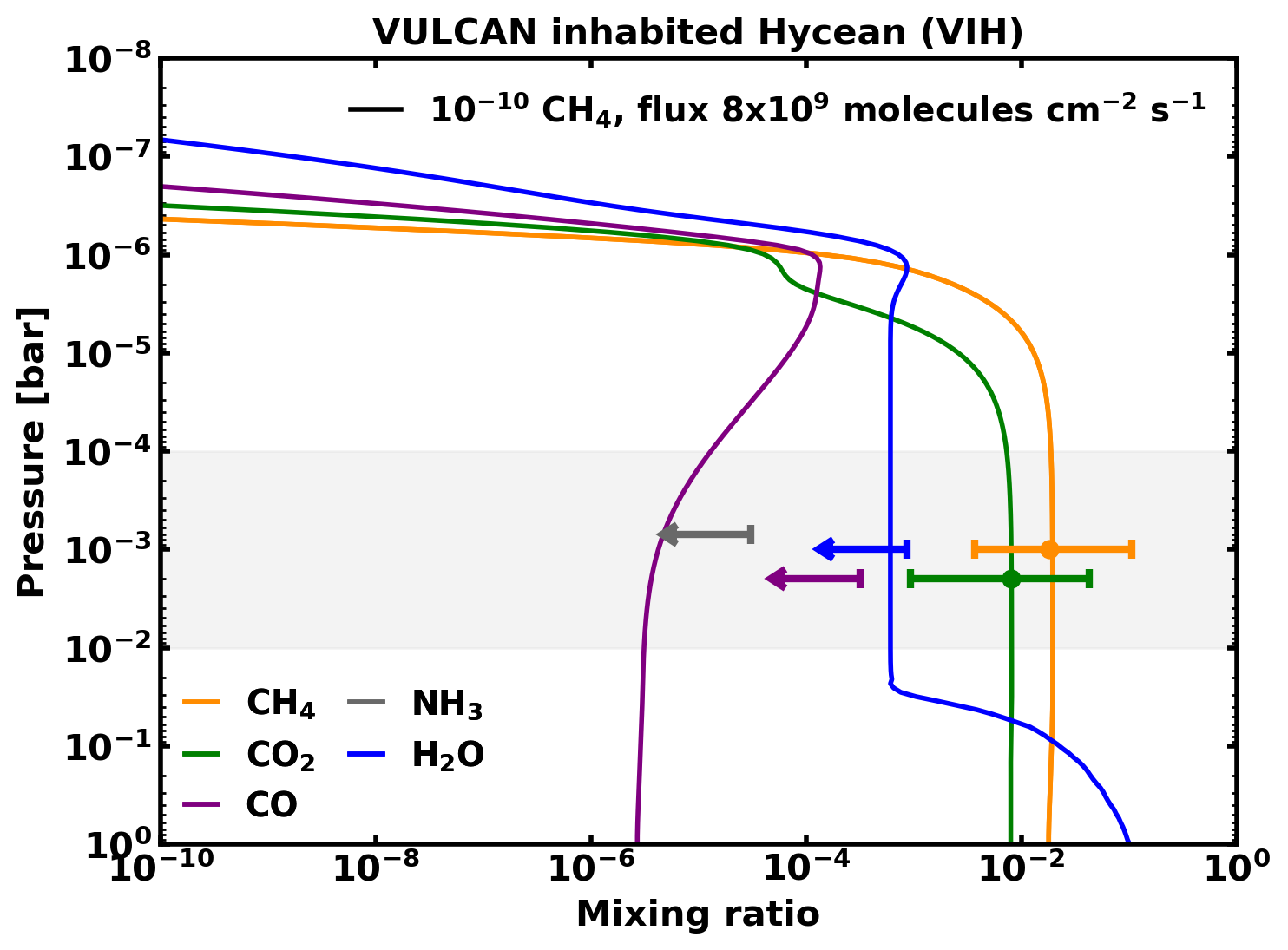}
    \caption{Inhabited Hycean scenario for K2-18~b using \textsf{VULCAN}. The GJ~176 spectrum with a 0.3 top of atmosphere albedo is used, as in W24, with the same $P$-$T$ and $K_\textrm{zz}$ profile from W24. The inhabited hycean scenarios imposes a flux for \ce{CH4} at the planetary surface of $8\times10^{9}$ molecules cm\textsuperscript{-2}. The inhabited simulation starts with a \ce{CH4} volume mixing ratio of $10^{-10}$. \ce{CO2} has a fixed mixing ratio condition at the surface of $8\times10^{-3}$. The mixing ratio of \ce{CH4}, \ce{CO2}, \ce{CO}, \ce{H2O}, and \ce{NH3} are plotted against atmospheric pressure. The constraints and colours are the same as in Fig.~\ref{Original Wogan cases figure}. This inhabited Hycean scenario is consistent with the retrieved abundance constraints from JWST observations \citep{2023ApJ...956L..13M} for all five molecules, similar to the W24 inhabited Hycean scenario.}
    \label{VULCAN inhabited Hycean figure}
\end{figure}

This result, that photochemical modeling of mini-Neptunes is incompatible with the retrieved JWST constraints, is also consistent with predicted abundances in \cite{2024ApJ...971L..48Y} and \cite{2024arXiv240709009H}.

\cite{2024ApJ...971L..48Y} simulated K2-18~b as a mini-Neptune using the \textsf{EPACRIS} model, exploring a parameter space in $P$-$T$ profiles, $K_\textrm{zz}$, and chemical composition. Their model was able to get compatible \ce{CH4} and \ce{CO2} abundances in their $100\times Z_0$ and $10\times Z_0$ cases with the retrieved ``two offset" constraints from \cite{2023ApJ...956L..13M}, and their $10\times Z_0$ case was also consistent with CO. However, \ce{NH3} and \ce{H2O} were more than an order of magnitude too high. Furthermore, considering the best fit ``one offset" constraints from \cite{2023ApJ...956L..13M}, CO is too high in their $10\times Z_0$ case. 

\cite{2024arXiv240709009H} simulated K2-18~b with and without water condensation, with albedos varying between 0 and 0.75, and surface pressures of 1 and 1000 bar. They concluded that the JWST observations were best explained by a deep mini-Neptune atmosphere with a 0.56 albedo with \ce{CH4} feature amplitudes greater than 100 ppm \citep[note that this contradicts the results from][]{2024A&A...686A.131L}. However, in this case, the predicted mixing ratios in the photosphere are $\approx 50$ and $\approx 2.5$ times higher than the $2$-$\sigma$ upper limits on \ce{NH3} and CO, respectively, and 3 times lower than the $1$-$\sigma$ limit for \ce{CO2}. The choice of 70 K for the internal temperature may be too high (see section \ref{K2-18 b planetary scenarios section}). Note that the corresponding shallow 1 bar case for K2-18~b in \cite{2024arXiv240709009H} does not explain the data either: \ce{H2O}, \ce{CO}, and \ce{NH3} are below the $2$-$\sigma$ upper limits, but \ce{CH4} and \ce{CO2} are several orders of magnitude under-abundant. Whilst \cite{2024arXiv240709009H} account for the albedo in setting the $P$-$T$ profile, they do not modify the total incoming UV flux whilst using an active M2V star at 45 Myr age, which may both result in an overestimation of the photolysis rates. This is important as it is photodissociation that leads to \ce{CH4} destruction in models of shallow Hycean atmospheres.

In order for the mini-Neptune scenario to explain the observed abundances for K2-18 b, it may require the following: a large enough Bond albedo (e.g., $\geq 0.4$) such that water sufficiently condenses below the photosphere; some mechanisms that can deplete both \ce{NH3} and CO, which may require a low C/O ratio of $\sim 0.02$ \citep{2023ApJ...956L..13M}; and have a higher internal heat flux than would otherwise be expected for its mass and age to produce enough \ce{CO2}.

\subsubsection{Uninhabited Hycean scenario}
\label{Uninhabited Hycean scenario section}

In W24, the \textsf{Photochem} uninhabited Hycean simulation starts with negligible \ce{CH4}, which then accumulates through photodissociation of \ce{CO2} and subsequent reactions. The \ce{CH4} abundance remains significantly lower than the observed abundance. We find that this result is the same even when the model starts with a high \ce{CH4} abundance by volume (e.g. 5\%). This result was used by W24 to argue that the uninhabited Hycean case scenario is incompatible with the observational data.

However, previous work has shown that \ce{CH4} can be retained at higher abundances when the surface pressure is greater due to thermochemical recycling \citep{2021ApJ...914...38Y, 2023FaDi..245...80M}. \cite{2023FaDi..245...80M} found that \ce{CH4} abundance was significantly depleted in a 1 bar versus a 100 bar Hycean case (e.g. volume mixing ratio of $\sim 10^{-5}$ versus $\sim 10^{-2}$). Motivated by this, we explore surface pressures between 1 and 100 bar. We also investigate the effect of the UV albedo of the planet on the \ce{CH4} abundance. We note that the UV albedo here does not correspond to the Bond albedo, but rather the scaling factor for the incoming stellar UV spectrum used, which was of GJ~436 in this instance. The motivation for this is the fact that Neptune itself is reported to have an albedo of up to 0.85 for UV and blue wavelengths of 300-500 nm \citep{2019Icar..331...69I, 2022JGRE..12707189I, 2023JGRE..12807980I}, but a total planetary Bond albedo of 0.3 \citep{1991JGR....9618921P}. We explore a range of UV albedos up to Neptune's value of 0.85 which reduces the photodissociation rates in the model. We use the PT3 $P$-$T$ profile from \cite{2023FaDi..245...80M}. The choice of PT3 is due to it having the most similar stratospheric temperature to the $P-T$ profiles used in W24 but extending to deeper pressures.

We show the results for a 25 bar and 50 bar surface pressure in Fig.~\ref{VULCAN uninhabited Hycean figure}a and Fig.~\ref{VULCAN uninhabited Hycean figure}b, respectively. The combination of the higher surface pressure and larger UV albedo results in a longer atmospheric lifetime for \ce{CH4}. We find that a case with a 50 bar surface pressure, with an initial metallicity of 100 times solar (starting with $\sim 5\%$ \ce{CH4} by volume), and a 0.75 UV albedo, can explain the abundances in the case of an uninhabited Hycean if sulfur is not included in the chemical network \citep[sulfur species may be deposited into the ocean on a Hycean world;][]{2021ApJ...921L...8H, 2019ApJ...887..231L, 2024ApJ...966L..24T}. When sulfur is included, \ce{CH4} is depleted but is still consistent with observations, but now the CO volume mixing ratio is significantly higher than the upper limit. With a 25 bar surface pressure, the predicted abundances provide a good fit for all five JWST constraints when initialising the model with $150$ times solar metallicity and invoking a UV albedo of 0.85. Again, CO is overabundant when including sulfur. In that scenario, \ce{OH} and H, which react with CO, are several orders of magnitude lower in abundance. Other species like S and \ce{H2S} can destroy CO and are only present when including sulfur.

The \ce{CH4} is continuing to deplete after $10^{17}$ s. This indicates that the lifetime of \ce{CH4} is on the order of or longer than $10^{17}$ s, which is more than the estimated age of K2-18 b \citep{2019RNAAS...3..189G}. As found previously, if there are uninhabited Hycean worlds which form with a large \ce{CH4} abundance (high metallicity), then older uninhabited Hyceans may exhibit depleted \ce{CH4}, whilst younger uninhabited Hyceans may have retained enough \ce{CH4} to be observable with JWST \citep{2023FaDi..245...80M}.

We tested other albedos, pressures, and metallicities. In several cases, \ce{CH4} matches the retrieved abundance but CO is too high. However, CO can dissolve in an ocean \citep{1973E&PSL..20..145V}, and a model of a prebiotic Archean Earth derived a CO deposition velocity of $10^{-8}$ cm s\textsuperscript{-1} \citep{2005Gbio....3...53K}. This deposition velocity depends on the partial pressure of CO, the overturning velocity of the ocean, the pH of the ocean, and the depth of the ocean, amongst other parameters \citep{2005Gbio....3...53K}. Such estimates are beyond the scope of this work because such an ocean on K2-18~b may span a wide parameter space \citep{2024MNRAS.529..409R}. Future work should investigate this in more detail, but it could be a viable mechanism whereby CO is adequately depleted.

In all simulations, due to the assumed presence of an ocean in which \ce{NH3} dissolves, \ce{NH3} is depleted to less than $10^{-10}$ at the surface. \ce{H2O} is below the 2-$\sigma$ upper limit due to cold trapping. If the temperature of the atmosphere near the cold trap was decreased or increased, then \ce{H2O} would correspondingly decrease or increase.

In contrast to W24, we find that an uninhabited Hycean scenario cannot be entirely ruled out by the data. A large number of factors effect the \ce{CH4} abundance, including the surface pressure, UV albedo, and the abundance of other molecules, such as sulfur bearing species, all of which have no observational constraints at present. The choice of initial conditions instead determines the final result \citep[as found previously in e.g.,][]{2021ApJ...921L...8H, 2023FaDi..245...80M}. We note that the plausibility of this scenario also depends on the temperature profile and whether a temperate surface ocean is actually physically possible for the assumed conditions.

Additionally, the destruction of \ce{CH4} is due to photodissociation, which does not occur on the nightside of a tidally locked exoplanet. \cite{2021ApJ...922L..27T} used the 2D version of \textsf{VULCAN} to simulate a 1 bar \ce{H2}-rich atmosphere. Without winds, \ce{CH4} was depleted on the dayside when compared to the nightside, homogenising to an intermediate value with winds. If this qualitative result holds for a variety of conditions, then multidimensional models may be required to explain how \ce{CH4} can match the abundances at the terminator in an uninhabited Hycean scenario if the UV albedo is not otherwise high enough to stop significant dayside \ce{CH4} destruction.

\subsubsection{Inhabited Hycean scenario}

The inhabited Hycean scenario assumes a biogenic source of CH$_4$ on K2-18~b, e.g. through methanogenesis, as well as the consumption of CO. Several studies have suggested the possibility of such a \ce{CH4} source on K2-18~b  \citep{2023FaDi..245...80M, 2023ApJ...956L..13M, 2024ApJ...963L...7W, 2024ApJ...964L..19G, 2024ApJ...966L..24T}. We recreate a similar experiment to the W24 inhabited Hycean case using \textsf{VULCAN} (instead of \textsf{Photochem} used by W24) and show the results in Fig.~\ref{VULCAN inhabited Hycean figure}. We use the same $P$-$T$ profile, stellar input spectrum, surface pressure, and fixed mixing ratios at the lower boundary. We initialise \ce{CH4} at a volume mixing ratio of $10^{-10}$.

To produce 2\% \ce{CH4} by volume in the atmospheric simulation, we find a flux of $8\times10^{9}$ molecules cm\textsuperscript{-2} s\textsuperscript{-1} using \textsf{VULCAN}, compared to the $5\times10^{10}$ molecules cm\textsuperscript{-2} s\textsuperscript{-1} found in W24. Considering that Earth's modern biogenic flux is $\sim 1\times10^{11}$ molecules cm\textsuperscript{-2} s\textsuperscript{-1} \citep{2020ERL....15g1002J}, and the surface of a Hycean world would be fully covered in an ocean, such a biogenic flux is not implausible.

Given a set of initial conditions, provided that the \ce{CH4} flux is sufficient to produce the observed \ce{CH4}, the inhabited Hycean scenario can be shown to be consistent with retrieved abundances. Alternatively, as shown in the previous section, an uninhabited Hycean atmosphere could also explain the CH$_4$ abundance under some circumstances. However, even if a biogenic source of CH$_4$ is not required to explain the data the possibility of life cannot be excluded. Moreover, the ``inhabited Hycean'' scenario can be thought of as an uninhabited Hycean which has an unknown source of \ce{CH4} and/or an unknown sink of CO.

\section{Summary and discussion}
\label{Summary and discussion section}

In this study we presented a detailed exploration of photochemical models to investigate different scenarios for the temperate sub-Neptune K2-18~b. Here we summarise our results and discuss the implications. We first provide a summary of the key results presented in this study and how they compare to the results in W24. Subsequently, we describe the implications of our work for future studies and observations of exoplanetary atmospheres.

\subsection{Summary}
\label{Summary section}

Following the photochemical modeling work of W24, we simulated the atmosphere of K2-18~b, assuming that it is either a Hycean world or a mini-Neptune with a deep atmosphere. With the results from \cite{rigby2024towards} that discussed the issues with the magma ocean scenario for K2-18~b \citep{2024ApJ...962L...8S}, and the lack of photochemical modelling in the supercritical ocean scenario \citep{2024arXiv240906258L}, we have assessed the remaining three proposed scenarios: uninhabited Hycean, inhabited Hycean, and mini-Neptune.

The mini-Neptune scenario with current modelling from this study and three different studies \citep{2024ApJ...963L...7W, 2024ApJ...971L..48Y, 2024arXiv240709009H} is unable to explain the JWST data because the predicted abundances of between 2--4 molecules are incompatible with the observed constraints. Depending on the initial assumptions, either CO or \ce{CH4}, or both molecules, are inconsistent with the abundance constraints in the uninhabited Hycean scenario. However, with a high enough surface pressure (e.g. 25 bar), a significant UV albedo of $\approx 0.85$ (and/or the case where the GJ~176, GJ~436, and GJ~849 spectra overestimate the UV flux of K2-18), then the uninhabited Hycean can explain the abundances, and the presence of a biosphere is not required in that scenario. Moreover, if the age of K2-18~b is on the lower end of the age estimate ($2.4 \pm 0.6$ Gyr), then the $\approx 3.2$ Gyr simulations we performed in \textsf{VULCAN} may have overestimated \ce{CH4} depletion. The inhabited Hycean scenario fits all five of the observational abundance constraints.

Whilst existing observations indicate there may be some clouds/hazes at the terminator for K2-18~b \citep{2023ApJ...956L..13M}, the albedo remains unconstrained, although the required planetary Bond albedo for a Hycean with a habitable ocean currently cannot be ruled out. Updated photochemical reaction networks, and 2D or 3D photochemical models, may produce mixing ratios that are compatible with observations for other scenarios. Future results from JWST observations are likely to narrow the constraints we have for the molecules in question (e.g., JWST C1 GO 2372; PI: Renyu Hu, JWST C1 GO 2722; PI: Nikku Madhusudhan).

Based on current observations and photochemical modelling, we now summarise the implications of our work:

\begin{itemize}
    \item In contrast to W24, we find that both uninhabited and inhabited Hycean scenarios  can explain the retrieved JWST abundances for K2-18~b given specific model assumptions. Future observations and photochemical simulations may favour a scenario different to a Hycean exoplanet, but existing observations and photochemical results currently support one of the Hycean scenarios. The mini-Neptune scenario of W24, using two independent 1D photochemical models, is inconsistent with retrieved mixing ratios over a range of initial conditions and sensitivity tests.
    \item Using inaccurate cross sections can modify the predicted abundance of key molecules (e.g. \ce{NH3, CO, CH4, SO2}) by factors of up to $\sim 10^{3}$ and $\sim~10^{9}$ for Hycean and mini-Neptune scenarios, respectively. Different assumed photolysis branching ratios, which were not tested over a large parameter space here, are also likely to alter chemical predictions. 
    \item Choosing only one stellar spectrum when modeling a planet with unknown incoming UV flux introduces further uncertainties, adding additional uncertainty when combined with different cross sections. We call for the spectrum of K2-18 to be measured with the Hubble Space Telescope. Until then, modellers should explore different stellar input spectra.
\end{itemize}

\subsection{Which star represents K2-18?}
\label{Discussion: Stellar spectra}

The UV spectrum of K2-18 is unknown. Photochemical models of K2-18~b have thus far used observed spectra from other stars that are seemingly representative of K2-18 due to comparable measured stellar properties (e.g. radius, $T_\textrm{eff}$, and rotation rate). In the Hycean and mini-Neptune simulations, we used three stellar spectra that, under the assumption of stellar proxies, can be substituted for the stellar spectrum of K2-18. These were the MUSCLES spectra for GJ~176, GJ~436, and GJ~849, which are given with fractional uncertainties of up to $\sim 10^{3}-10^{4}$ for the observed flux density. Out of all available spectra, it is possible that either GJ~176, or GJ~436, or GJ~849, does best represent K2-18. But this is not a certainty, and thus represents a general problem in accurately modeling exoplanet atmospheres. Furthermore, how the star and planetary atmosphere evolves through time is an important question regarding the current atmospheres we are able to observe with JWST.

Then there is the separate problem of stellar activity and flares, which are known to impact photochemistry in terrestrial atmospheres. When modeling terrestrial exoplanet atmospheres using a 3D model, \cite{2021NatAs...5..298C} found that frequent flaring reduces \ce{O3} on K and M dwarf exoplanets, whilst \cite{2023MNRAS.518.2472R} found the opposite for a terrestrial Proxima Centauri~b: flaring increased atmospheric \ce{O3} by 20 times. Note that many of the assumptions in these two models were not the same. However, this does raise the question of whether such opposite predictions could occur for photochemically active species in temperate sub-Neptune atmospheres when using different photochemical models. The impact of flares on Hycean atmospheres remains untested and we leave that for future investigations.

\subsection{Cross sections and branching ratios}
\label{Results: Cross sections and branching ratios}

Cross sections and branching ratios are key inputs to photochemical models which influence the vertical profile of many chemical species. As demonstrated by W24, a seemingly small change to a branching ratio for \ce{H2O} photolysis at a single wavelength (Lyman-$\alpha$), can result in a big change for a specific molecule, like \ce{CH4}. Similarly here, Fig.~\ref{Branching ratio figure} shows large differences in branching ratios between models for multiple key species, and alongside discrepant cross sections (see Fig.~\ref{All XS figure}), this may result in further uncertainties depending on the model and the chemical network. 

As an initial test, we swapped the branching ratios from \textsf{VULCAN} into the standard W24 mini-Neptune setup (GJ~176, M/H = 100) for \ce{CO2}, \ce{CH4}, \ce{NH3}, \ce{SO2}, and \ce{O3} and performed simulations with the \textsf{Photochem} and \textsf{VULCAN} binned and native cross sections. The largest perturbation observed in the photosphere was in the \textsf{VULCAN} native cross section case, where \ce{NH3} and \ce{SO2} decreased in the photosphere by up to a factor of 1100 and 900, respectively. 

In simulations of terrestrial planetary atmospheres, the extended \ce{H2O} cross sections, as recommended in \cite{2020ApJ...896..148R}, impacted \ce{H2O} mixing ratios and that of OH. Subsequently, \cite{2024ApJ...967..114B} showed that for anoxic (lack of oxygen) terrestrial exoplanets orbiting FGKM host stars, the extended cross sections were the least important for M dwarf stars. Because \textsf{Photochem} does not use the recommended cross sections from \cite{2020ApJ...896..148R}, this implies that the results we present may have even greater implications for Hycean and mini-Neptune exoplanets orbiting FGK stars, but that hypothesis must be tested. 

Alternative chemical networks to those used here may change the results and conclusions of our work. For instance, there may be a mechanism which is able to replenish \ce{CH4} in a Hycean atmosphere, or where CO and \ce{NH3} are depleted in a mini-Neptune atmosphere.

Awareness of the impact of different assumptions and chemical input data sources is vital; having accurate and updated chemical input data is important for future interpretation of planetary spectra, and may influence whether a biosignature is attributed to a biogenic or abiogenic source.

\subsection{Implications for other planetary interpretation}
\label{Implications for other planetary interpretation}

It is important to consider how photochemical codes respond to different input data and initial conditions, because this can affect interpretations of other planetary spectra beyond K2-18~b. Recently, TOI-270~d was observed by JWST and the observations were analysed by two different groups \citep[][GO Program 4098]{2024arXiv240303325B, 2024A&A...683L...2H}. Both analyses found evidence for \ce{CO2} and \ce{CH4} similar in abundance to K2-18~b, evidence for \ce{H2O}, and weak evidence for \ce{CS2}, a possible biosignature \citep{2013ApJ...777...95S, 2018AsBio..18..663S}. \cite{2024arXiv240303325B} also report a possible signature of \ce{SO2}, which \cite{2024A&A...683L...2H} did not find. Neither studies detected \ce{NH3} or \ce{CO}, similar to the K2-18~b observations. 

\cite{2024A&A...683L...2H} report that the simultaneous detection of \ce{CH4} and \ce{CO2}, without the detection of \ce{NH3}, lends weight to the argument that TOI-270~d could be a Hycean exoplanet. On the other hand, \cite{2024arXiv240303325B} argued that their abundance constraints for the atmosphere of  TOI-270~d can be explained by a ``Miscible-Envelope Sub-Neptune" exoplanet. To support their argument, \cite{2024arXiv240303325B} adapted the mini-Neptune setup of W24 to TOI-270~d, assuming a metallicity of $230 \times$ solar and retaining the stellar spectrum to be of GJ~176. They claim the model abundances match the retrieved constraints for \ce{H2O}, \ce{CO}, \ce{CO2}, \ce{NH3}, and \ce{CH4}. However, we find that those abundances are incompatible with the retrieved abundance constraints of \cite{2024A&A...683L...2H} for \ce{CO}, \ce{CH4}, and \ce{H2O}, for the ``One offset + DT" case. Secondly, including different cross sections for TOI-270~d as we have here would affect \ce{NH3} photolysis and impact its observability, because the pressure at which the \ce{NH3} mixing ratio starts to decrease can change by a factor of $\sim 100$. \cite{2024arXiv240303325B} find weak evidence for \ce{SO2} in the atmosphere of TOI-270~d, but don't show the \ce{SO2} mixing ratio in their figure 10 when simulating TOI-270~d with \textsf{Photochem}. Again, using different cross sections would change their predicted \ce{SO2} mixing ratio. 

\cite{2024arXiv240303325B} state that their atmospheric simulation of TOI-270~d using the model \textsf{EPACRIS} \citep{2024ApJ...966..189Y} does produce the right \ce{CO2}, \ce{H2O}, \ce{CH4}, and \ce{SO2} mixing ratios, but it is unclear if \textsf{EPACRIS} predicts concentrations for other molecules (\ce{CO} and \ce{NH3}) that are consistent with the retrieved constraints as those results are not reported. Their modeling finds much lower \ce{CS2} mixing ratios ($10^{-10}$ near 1 mbar) than the tentatively detected abundance of \ce{CS2} ($\approx 3\times10^{-4}$ at $2.55 \sigma$). Finally, it is arguable that the star GJ~163 better represents the host star TOI 270 more than GJ~176 (see Appendix \ref{Appendix: The assumed stellar spectrum} and Table \ref{Star spectral type table}); using GJ~163 may influence the results presented in \cite{2024arXiv240303325B}.

\subsection{Tidally locked anisotropy}
\label{Discussion: 3D and tidally locked anisotropy}

When exoplanets pass in front of their host stars, transmission spectra of their atmospheres can be taken to reveal absorption signatures from any chemical species present. Transmission spectra probe the terminator region of an exoplanet's atmosphere. Photochemical modeling in 1D may not adequately capture this portion of the atmosphere. 3D modeling of terrestrial exoplanets that are tidally locked have shown that photochemically active molecules can be distributed inhomogeneously \citep{2016EP&S...68...96P, 2023ApJ...959...45C, 2023MNRAS.526..263B}, including at orbital periods longer than K2-18~b \citep{2018ApJ...868L...6C, 2019ApJ...886...16C}, where DMS showed large day-to-nightside anisotropy \citep{2018ApJ...868L...6C}. Similarly for larger exoplanets such as hot Jupiters, 2D and 3D models are critical for understanding chemical anisotropy and explaining observed spectra \citep{2019ApJ...880...14S, 2022A&A...664A..82T, 2023A&A...672A.110L, 2023ApJ...951..117S, 2023Natur.617..483T, 2024ApJ...963...41T, 2024Natur.632.1017E}. There is no reason to expect that temperate sub-Neptunes, including mini-Neptunes and Hycean exoplanets, would be any different. 

Our results in section \ref{Uninhabited Hycean scenario section} regarding an uninhabited Hycean scenario showed that pressures of tens of bar and an adequate UV albedo (e.g. 0.75) may be necessary to prevent substantial \ce{CH4} depletion below the abundance constraint for K2-18~b ($\sim 1\%$ by volume). Nevertheless, a 25 bar Hycean world with a habitable ocean could be physically infeasible due to a steam runaway greenhouse \citep{2020ApJ...898...44S, 2023ApJ...953..168I, 2023ApJ...944...20P, 2024A&A...686A.131L}. The abundance of \ce{CH4} in a case with a shallower ocean surface (e.g., 10 bar) will be dependent on many free model parameters that are not constrained, which include: the UV flux of the host star, the albedo (both Bond and UV), the efficiency of \ce{CH4} recycling on the nightside of a tidally locked exoplanet, and the age of K2-18~b. Further explorations into such a parameter space demand the use of multi-dimensional models to comprehensively assess the efficacy of the uninhabited Hycean scenario for K2-18~b, and other Hycean candidates such as TOI-270~d.

\subsection{Methane as a biosignature}

In the context of a potentially habitable exoplanet, a molecule present in the exoplanet's atmosphere that indicates biological activity can be considered a biosignature \citep{2017PhR...713....1G, 2018AsBio..18..630M, 2024RvMG...90..465S}.

\ce{CH4} is a molecule on Earth that has both abiotic and biotic sources. In an atmosphere composed primarily of reducing gasses (e.g. \ce{H2}), similar to the weakly reducing conditions expected during the Archean eon (4 - 2.4 Gyr ago) on Earth \citep{2001Sci...293..839C, 2005PreR..137..119K, 2020SciA....6.1420C}, methanogenesis could add significant quantities of \ce{CH4} to the atmosphere.

On an Earth-like rocky exoplanet, the atmospheric combination of \ce{CO2} and \ce{CH4}, with a lack of CO, is thought to be a possible biosignature \citep{2018SciA....4.5747K, 2019BAAS...51c.158K}. False positives of this combination in rocky exoplanetary atmospheres may be improbable \citep{2020PSJ.....1...58W, 2022PNAS..11917933T}. However, on rocky exoplanets with \ce{H2} dominated atmospheres, \ce{CH4} is not an unambiguous sign of life because it can be produced through photochemistry or geochemistry \citep{2013ApJ...777...95S}. Such geochemical sources consist of serpentinizing reactions and magmatic outgassing \citep{2022PNAS..11917933T}, and these processes are not expected to occur on ocean and Hycean worlds \citep{2009ApJ...700.1732K, 2018ApJ...864...75K, 2023FaDi..245...80M}.

Given all the simulations here, and those that exist within the literature, the inhabited Hycean fits the observations best, followed by the uninhabited Hycean, while the mini-Neptune is more difficult to reconcile with observations. It is therefore pertinent to consider whether the observed \ce{CH4} in the atmosphere of K2-18~b might be from a biological source.

Previous studies have argued that \ce{CH4} could be produced by extraterrestrial life on Hycean worlds \citep{2023FaDi..245...80M, 2023ApJ...956L..13M, 2024ApJ...963L...7W, 2024ApJ...964L..19G, 2024ApJ...966L..24T} and thus considered a potential biosignature in the context of a habitable environment. Despite this, the relative proportions of possible biotic versus abiotic sources in a Hycean environment is not well understood and requires further exploration. If K2-18~b is a Hycean exoplanet, whether it is viable to disentangle the ambiguity of the \ce{CH4} source remains an open question, and other lines of evidence will likely be necessary to determine if the planet is inhabited, rather than just habitable.

Overall, the interpretation of atmospheric observations using photochemical models of an exoplanet atmosphere depends on various factors, including the initial conditions, the boundary conditions, the number of dimensions included, and the input parameters such as chemical reaction rates, photochemical cross sections, and the assumed stellar spectrum. Therefore, conclusions drawn from JWST exoplanet atmospheric observations should be aware of the many uncertainties that exist in photochemical codes (e.g. cross sections), stellar spectrum inputs (MUSCLES observations and synthetic spectra), and be cautious when not including 3D effects, because transmission spectra probe the exoplanet's terminator. Such critical assessment is necessary now that astronomers have started to observe potentially habitable exoplanet atmospheres. 

\section*{Acknowledgments}

We thank the anonymous reviewers for their valuable comments. We thank Nick Wogan, the \textsf{Atmos} team, and Shami Tsai, for making their models \textsf{Photochem}, \textsf{Atmos}, and \textsf{VULCAN}, respectively, publicly available on GitHub. GC thanks Nicholas Wogan and Shami Tsai for useful advice regarding modeling with \textsf{VULCAN}, Lalitha Sairam and Måns Holmberg for helpful discussions, and Savvas Constantinou for advice and guidance with modeling.


\bibliography{references}{}
\bibliographystyle{aasjournal}

\setcounter{table}{0} 
\renewcommand{\thetable}{A\arabic{table}}

\setcounter{figure}{0} 
\renewcommand{\thefigure}{A\arabic{figure}}

\appendix

\section{Simulations performed}
\label{Appendix: Simulations performed}

We performed hundreds of simulations over the three scenarios, using both the W24 setup and the \textsf{VULCAN} 1D photochemical model. The simulations using the W24 setup are shown in Table~\ref{Boundary condition table W24 setup}, and those using \textsf{VULCAN} are shown in Table~\ref{Boundary condition table VULCAN}. 
We altered cross section sources and their wavelength resolution, as well as the stellar input spectrum in the W24 setup. See Appendix \ref{Appendix: Cross sections and branching ratios} for more information regarding photochemical cross sections and branching ratios. The three different stellar spectra used in the W24 setup were GJ~176, GJ~436, and GJ~849. GJ~176 and GJ~436 were used in \textsf{VULCAN} simulations. Appendix \ref{Appendix: The assumed stellar spectrum} details why these stars were chosen. 

The boundary conditions are given in Table~\ref{Boundary condition table W24 setup} and Table~\ref{Boundary condition table VULCAN}. If a boundary condition is not listed in these tables, then the molecule has a zero flux boundary condition. An exception to this is H$_2$O, the abundance profile of which is based on the humidity and the $P$-$T$ profile.

\begin{table}[b!]
\centering
\caption{The boundary conditions and inputs for simulations that used the W24 setup are shown. This includes the volume mixing ratios and fluxes of species at the lower boundary, the initial metallicity, the vertical deposition of various chemical species, the eddy diffusion parameter ($K_\textrm{zz}$), the cross section ($\sigma_\lambda$) source, the resolution of the cross section source (native or binned), the stellar input spectrum, and the top of atmosphere albedo. Text in bold are the original parameters that were included in the final simulation results shown in W24, and we note that W24 did test a wider range of parameters that the ones they presented.}
\label{Boundary condition table W24 setup}
\begin{tabular}{@{}lccc@{}}
\toprule
Parameter & Uninhabited Hycean & Inhabited Hycean & Mini-Neptune \\ \midrule
\ce{H2O} humidity & \textbf{1.0} & \textbf{1.0} & \textbf{1.0} \\
\ce{N2} mixing ratio & $\mathbf{3\times10^{-3}}$ & $\mathbf{3\times10^{-3}}$ & \textbf{See metallicity} \\
\ce{CO2} mixing ratio & $\mathbf{8\times10^{-3}}$ & $\mathbf{8\times10^{-3}}$ & \textbf{See metallicity} \\
\ce{CH4} flux [cm\textsuperscript{-2} s\textsuperscript{-1}] & -- & $\mathbf{5\times10^{10}}$ & -- \\
CO $v_\textrm{dep}$ [cm s\textsuperscript{-1}] & -- & $\mathbf{1.2\times10^{-4}}$ & -- \\
HCCCN $v_\textrm{dep}$ [cm s\textsuperscript{-1}] & $\mathbf{7\times10^{-3}}$ & $\mathbf{7\times10^{-3}}$ & $\mathbf{K_{zz} / H}$ \\
HCN $v_\textrm{dep}$ [cm s\textsuperscript{-1}] & $\mathbf{7\times10^{-3}}$ & $\mathbf{7\times10^{-3}}$ & -- \\
\ce{NH3} $v_\textrm{dep}$ [cm s\textsuperscript{-1}] & -- & -- & -- \\
\ce{C2H6} $v_\textrm{dep}$ [cm s\textsuperscript{-1}] & $\mathbf{10^{-5}}$ & $\mathbf{10^{-5}}$ & $\mathbf{K_{zz} / H}$ \\
Metallicity ($\times$ solar) & -- & -- & 30, 50, \textbf{100}, 200 \\
C/O ratio ($\times$ solar) ($\times$ solar) & -- & -- & 0.5, \textbf{1.0}, 2.0 \\
$T_\textrm{int}$ [K] & -- & -- & 30, 40, 50, \textbf{60}, 70 \\
$K_\textrm{zz}$ [cm\textsuperscript{2} s\textsuperscript{-1}] &  $\mathbf{5\times10^{5}}$ & $\mathbf{5\times10^{5}}$ & \textbf{See W24 figure A1b} \\
$\sigma_\lambda$ source (resolution) & \multicolumn{3}{@{}c@{}}{\textsf{\textbf{Photochem} (N)}, \textsf{VULCAN} (N, B), \textsf{Atmos} (N, B)} \\
Stellar spectrum &
\multicolumn{3}{@{}c@{}}{\textbf{GJ~176}, GJ~436, GJ~849} \\
Top of atmosphere albedo & \textbf{0.3} & \textbf{0.3} & \textbf{0.0}, 0.3, 0.4 \\
Model equilibrium time & $\mathbf{10^{14}}$ s & $\mathbf{10^{15}}$ s & $\mathbf{10^{10}}$ s \\
\bottomrule
\end{tabular}
\end{table}

For the W24 Hycean cases, we do not vary $K_\textrm{zz}$, the \ce{N2} concentrations (apart from when metallicity is specified and \ce{N2} is set by chemical equilibrium), or the tropospheric relative humidity, which is kept at 1. This is because W24 described how their uninhabited Hycean scenario result was insensitive to these specified parameters. In addition, the inhabited case is consistent over a variety of parameters, because it just requires tuning the \ce{CH4} surface-to-atmosphere flux to get a specific \ce{CH4} mixing ratio. Should K2-18~b be a Hycean exoplanet, given the potential large volume and surface coverage of an ocean relative to the Earth's ocean \citep{2024MNRAS.529..409R}, higher \ce{CH4} fluxes than those present on Earth could be possible, so this is not unfounded in the inhabited Hycean scenarios.
 
For the mini-Neptune cases using the W24 setup, we vary metallicity between 30 -- $200\times$ solar metallicity, the stellar spectrum, the cross section source, the cross section resolution with wavelength, the planetary internal temperature, and the C/O ratio between 0.5 -- $2\times$ solar. For a small number of cases shown in Fig.~\ref{Mini-Neptune albedo figure}, we varied the Bond albedo.

We perform only one \textsf{VULCAN} inhabited Hycean simulation to show that the model is consistent with all five abundance constraints. In the \textsf{VULCAN} uninhabited Hycean simulations, we vary initial metallicity, the stellar spectrum, the photodissoication albedo (not the Bond albedo), and the surface pressure. We also tested two photochemical networks.

\begin{table}[t!]
\centering
\caption{Photochemical simulations performed with \textsf{VULCAN} for K2-18~b. All simulations used either the ``SNCHO\textunderscore photo\textunderscore network\textunderscore 2024.txt'' or the ``NCHO\textunderscore photo\textunderscore network\textunderscore .txt'' chemical network. The network without sulfur was used in the Hycean cases, and with sulfur in the mini-Neptune cases, unless otherwise specified in the caption/legend of the figure. Simulations were performed with GJ~176 for the 1 bar Hycean cases, and with a 0.3 top of atmosphere albedo, as used in W24. At higher pressures, we used GJ~436. Cases either start with an initial \ce{CH4} mixing ratio ($f_{\ce{CH4}}$) which is specified arbitrarily, or a \ce{CH4} mixing ratio resulting from initial chemical equilibrium conditions set by the metallicity and the $P$-$T$ profile, which was the same as the one used in the W24 Hycean cases for 1 bar. At higher pressures, we used the PT3 profile from \cite{2023FaDi..245...80M}. The mixing ratio of ($f_{\ce{CO2}}$) is fixed at the lower boundary at the value specified. All simulations use a deposition velocity at the lower boundary to simulate \ce{NH3} dissolving in an ocean, and the dry deposition of HCN, \ce{C2H6}, and in the inhabited Hycean case only, CO. Text in bold are the original parameters that were included in the final simulation results shown in W24.}
\label{Boundary condition table VULCAN}
\begin{tabular}{@{}cccccccc@{}}
\toprule
Parameter & Uninhabited Hycean & Inhabited Hycean & Mini-Neptune \\ \midrule
\ce{H2O} humidity & 0.5, \textbf{1.0} & \textbf{1.0} & \textbf{1.0} \\
\ce{N2} mixing ratio & $\mathbf{3\times10^{-3}}$ & $\mathbf{3\times10^{-3}}$ & \textbf{See metallicity} \\
\ce{CO2} mixing ratio & $1,2\times10^{-3}$, $\mathbf{8\times10^{-3}}$ & $\mathbf{8\times10^{-3}}$ & \textbf{See metallicity} \\
\ce{CH4} flux [cm\textsuperscript{-2} s\textsuperscript{-1}] & -- & $8\times10^{10}$ & -- \\
CO $v_\textrm{dep}$ [cm s\textsuperscript{-1}] & -- & $\mathbf{1.2\times10^{-4}}$ & -- \\
HCCCN $v_\textrm{dep}$ [cm s\textsuperscript{-1}] & $\mathbf{7\times10^{-3}}$ & $\mathbf{7\times10^{-3}}$ & $\mathbf{K_{zz} / H}$ \\
HCN $v_\textrm{dep}$ [cm s\textsuperscript{-1}] & $\mathbf{7\times10^{-3}}$ & $\mathbf{7\times10^{-3}}$ & --- \\
\ce{NH3} $v_\textrm{dep}$ [cm s\textsuperscript{-1}] & -- & -- & -- \\
\ce{C2H6} $v_\textrm{dep}$ [cm s\textsuperscript{-1}] & $\mathbf{10^{-5}}$ & $\mathbf{10^{-5}}$ & $\mathbf{K_{zz} / H}$ \\
Metallicity ($\times$ solar) & 100--200 & -- & 50, \textbf{100}, 200 \\
C/O ratio ($\times$ solar) & 1.0 & -- & \textbf{1.0}\\
$T_\textrm{int}$ [K] & -- & -- & \textbf{60} \\
$K_\textrm{zz}$ [cm\textsuperscript{2} s\textsuperscript{-1}] &  $\mathbf{5\times10^{5}}$ & $\mathbf{5\times10^{5}}$ & \textbf{See W24 figure A1b} \\
$\sigma_\lambda$ source (resolution) & \multicolumn{3}{@{}c@{}}{ \textsf{VULCAN} (N)} \\
Stellar spectrum &
\multicolumn{3}{@{}c@{}}{\textbf{GJ~176}, GJ~436} \\
Top of atmosphere albedo & \textbf{0.3} & \textbf{0.3} & \textbf{0.0} \\
Model equilibrium time & $\mathbf{10^{17}}$ s & $\mathbf{10^{17}}$ s & Steady state criterion \\
\bottomrule
\end{tabular}
\end{table}

\section{Cross sections and branching ratios}
\label{Appendix: Cross sections and branching ratios}

\setcounter{table}{0} 
\renewcommand{\thetable}{B\arabic{table}}

\setcounter{figure}{0} 
\renewcommand{\thefigure}{B\arabic{figure}}

Photodissociation cross sections and their branching ratios form part of the critical input data for photochemical models. They are crucial for robust calculations of radiative transfer and chemical reactions in planetary atmospheres. We further describe the cross section and branching ratios present in the three models we sourced cross sections from.

\textsf{Photochem} cross section citations are available in a metadata.yaml file\footnote{\href{https://github.com/Nicholaswogan/photochem\_clima\_data/blob/20e95899182cdb336d5b45c98f7055d680417e42/xsections/metadata.yaml}{Photochem github cross section metadata.yaml}}, but the original source of many of these cross sections are unknown. The majority of the \textsf{VULCAN} cross sections come from the Leiden database\footnote{\href{http://home.strw.leidenuniv.nl/~ewine/photo}{http://home.strw.leidenuniv.nl/~ewine/photo}}, with a few others sourced from the PhiDrates\footnote{\href{https://phidrates.space.swri.edu/}{https://phidrates.space.swri.edu/}} or MPI-Mainz UV/VIS Spectral Atlas\footnote{\href{https://www.uv-vis-spectral-atlas-mainz.org/uvvis/}{https://www.uv-vis-spectral-atlas-mainz.org/uvvis/}} databases. Cross sections references for \textsf{Atmos} come from a variety of sources and some of those sources are given in the relevant folders on GitHub\footnote{\href{https://github.com/VirtualPlanetaryLaboratory/atmos/tree/master/PHOTOCHEM/DATA/XSECTIONS/}{Atmos Github cross section folder}}. These include MPI-Mainz, JPL recommendations, and some from unknown sources.

\begin{figure*}[t!]
	\centering
	\includegraphics[width=1\textwidth]{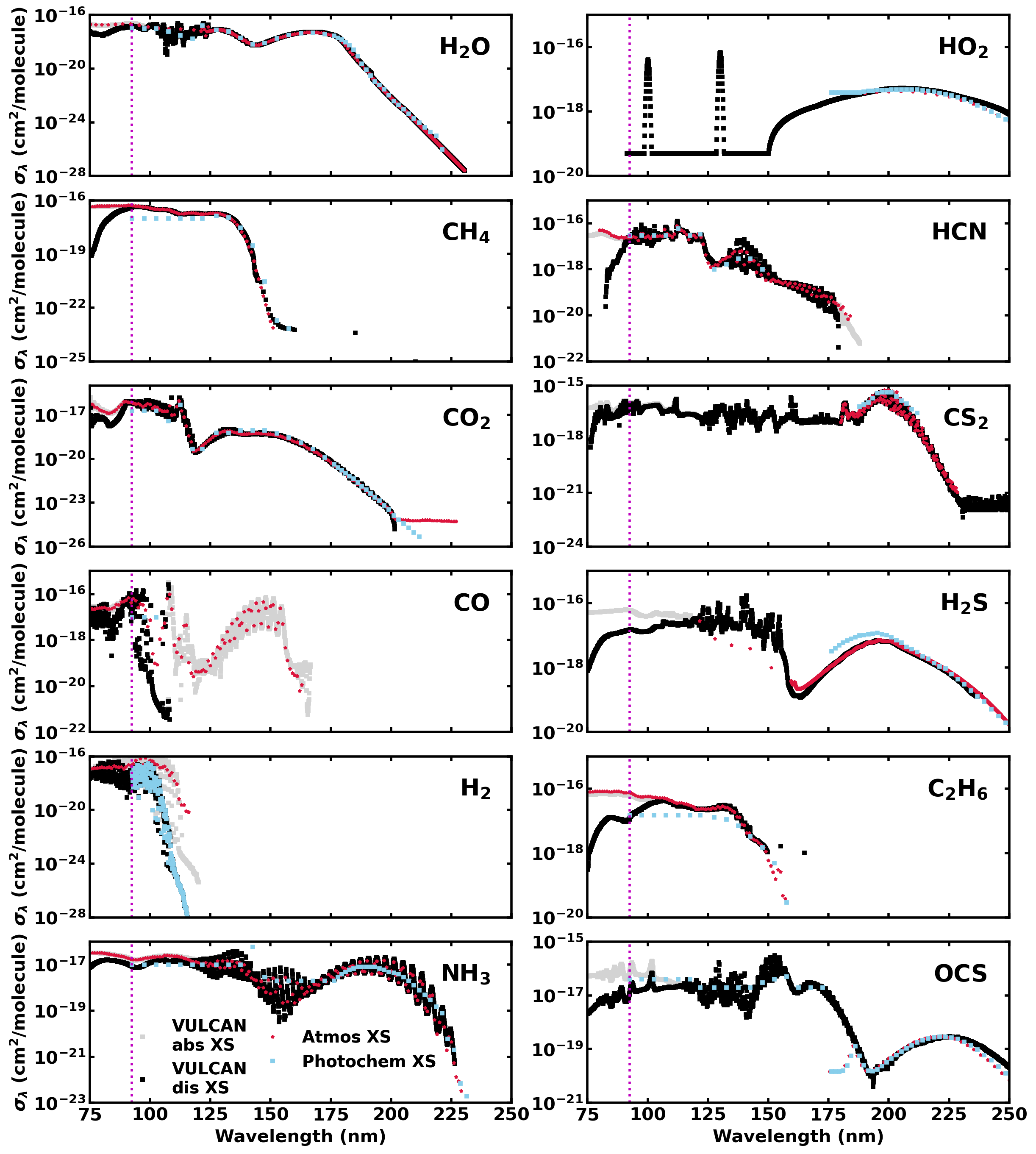}
    \caption{Photodissociation cross sections for twelve molecules are compared across three models. Cross sections ($\sigma_\lambda$) in terms of cm\textsuperscript{2} molecule\textsuperscript{-1} are shown for \ce{H2O}, \ce{CH4}, \ce{H2}, \ce{CO}, \ce{CO2}, \ce{NH3} in the left column on a log scale and the right column shows the cross section for \ce{HO2}, \ce{HCN}, \ce{CS2}, \ce{H2S}, \ce{C2H6}, and OCS and linear (right) scale versus wavelength in nm. Available cross sections in \textsf{Atmos} and \textsf{Photochem} are shown in red and light blue, respectively. \textsf{VULCAN} has cross sections given in terms of absorption (grey) and dissociation (black). The vertical magenta dotted line shows where the \textsf{Photochem} photolysis grid starts at 92.5 nm.} 
    \label{All XS figure}
\end{figure*}

We noticed that the photodissociation cross sections ($\sigma_\lambda$) for several species in \textsf{Photochem} do not closely match up-to-date measurements. We found that the data for important atmospheric molecules, such as \ce{NH3}, \ce{CO}, \ce{CO2}, \ce{CH4}, \ce{H2O}, and \ce{H2}, had significant differences. These are all shown in Fig.~\ref{All XS figure}, alongside the cross sections for \ce{HO2}, \ce{HCN}, \ce{CS2}, \ce{H2S}, \ce{C2H6}, and OCS. Depending on whether the \textsf{VULCAN} photoabsorption or photodissociation cross sections are considered, the data may show similarities or discrepancies with the \textsf{Atmos} cross sections. 

We now describe a few important molecules and how their cross sections vary between models. The \ce{H2O} cross section data is very similar between all three models from 125 -- 220 nm. However, longward of 221 nm, it appears that the \ce{H2O} cross sections have not been extended in \textsf{Photochem} as recommended by \cite{2020ApJ...896..148R}, with these wavelengths important for \ce{OH} production. Between 92.5 -- 127.5 nm, \ce{CH4} cross sections for \textsf{Photochem} have smaller values compared to the other two models. The  \ce{CH4} cross sections end at 151.0 nm and 157.5 nm in \textsf{Atmos} and \textsf{Photochem}, respectively, but continue in \textsf{VULCAN} until 237.3 nm. All three models significantly diverge for \ce{CO2} longward of 200 nm and this will affect production rates for CO and O (and also is impacted by the specified \ce{CO2} mixing ratio) in atmospheric modeling. The $\sigma_\lambda$ data for \ce{CO} in \textsf{Photochem} has the same cross section (1.00000000e-17) for three given wavelengths (92.5 nm, 97.5 nm, 102.5 nm), and 0 at all other given wavelengths, whereas \textsf{Atmos} and \textsf{VULCAN} have \ce{CO} cross sections varying over 5 orders of magnitude between 92 nm and 160 nm. \ce{H2}, which makes up the majority of these mini-Neptune and Hycean atmospheres, has large cross section differences in magnitude and shape between all three models between 90 and 120 nm. In the case of \ce{NH3}, between 92.5 - 142.5 nm, all three models have varied $\sigma_\lambda$ values which do not closely match. After 142.5 nm, differences arise due to the assumed resolution of the data.

\begin{figure*}[t!]
	\centering
	\includegraphics[width=1\textwidth]{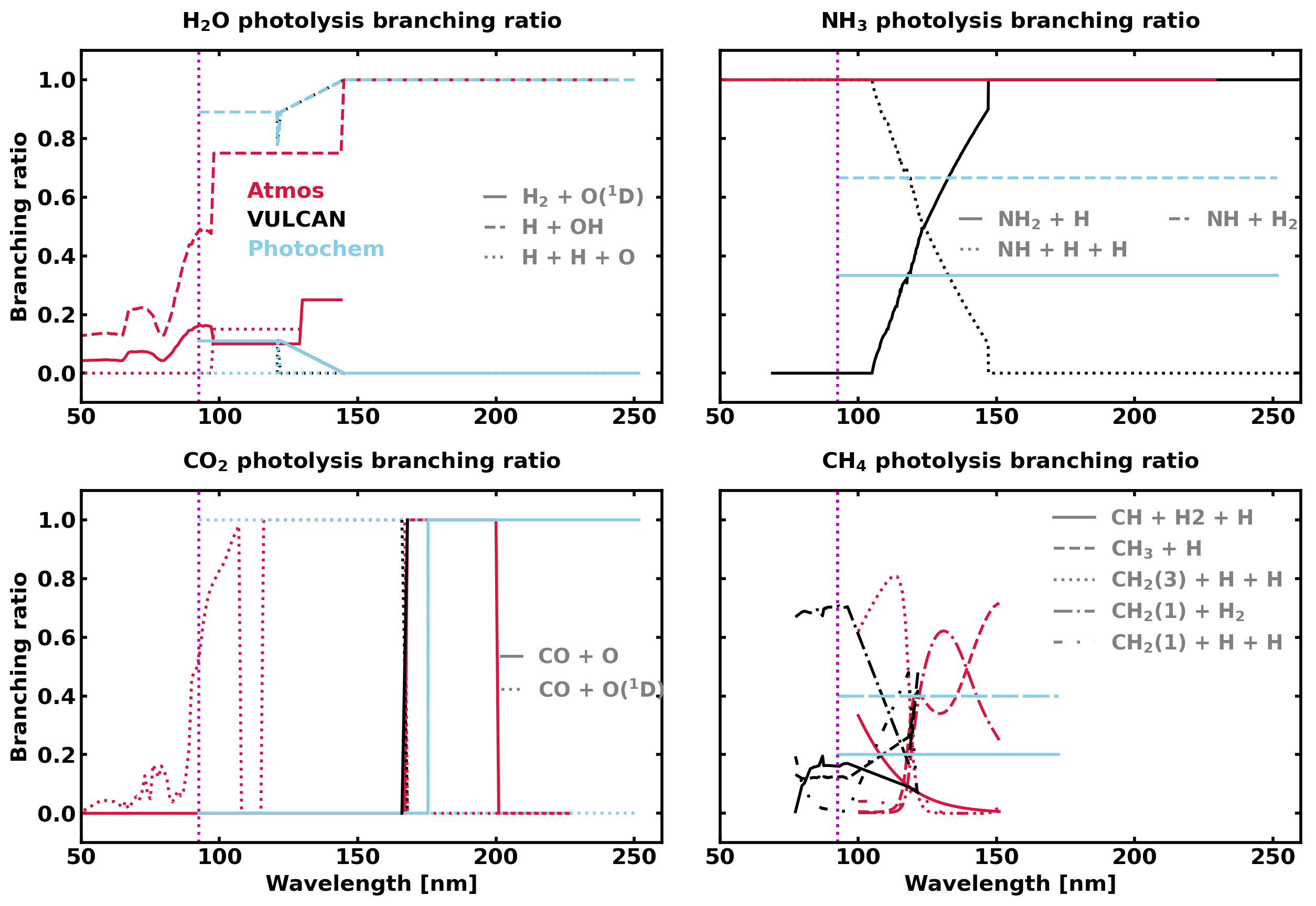}
    \caption{Branching ratios for four molecules compared between three models. Photolysis channels and their respective branching ratios are shown for \ce{H2O}, \ce{NH3}, \ce{CO2}, \ce{CH4} versus wavelength in nm. The data are taken from \textsf{VULCAN}, \textsf{Atmos}, and \textsf{Photochem}, and are shown in black, red, and light blue, respectively. \ce{H2O} branching ratio data between \textsf{Photochem} and \textsf{VULCAN} match, but differ for \textsf{Atmos}. The data for the three other molecules does not match. Additionally, not all three models consider the same photolysis channels.} 
    \label{Branching ratio figure}
\end{figure*}

Whilst not shown in Fig.~\ref{All XS figure}, the cross sections for \ce{N2}, \ce{N2O}, \ce{NO2}, \ce{SO}, \ce{SO2},  \ce{C2H2}, \ce{C2H4}, \ce{H2CO}, \ce{HNO3}, and \ce{H2O2}, also exhibit large discrepancies between the three models at wavelengths relevant to the \textsf{Photochem} photolysis grid. This is 22 molecules in total. Discrepancies also exist for \ce{CH2CO}, \ce{HS}, \ce{CH3CN}, \ce{CH3OH}, \ce{OH}, \ce{NO}, and \ce{H2SO4}, but these are not available in all three models, so we do not swap these molecules. Furthermore, \textsf{Photochem} does not include cross sections for \ce{CH3}, \ce{CH2}, or \ce{CH}, which do undergo photodissociation at UV wavelengths. Note that the resolution of the cross section ($\sigma_\lambda$) data for each model and molecule varies.

We tested simulations where we only swapped 6 (\ce{NH3}, \ce{CO}, \ce{CO2}, \ce{CH4}, \ce{H2O}, and \ce{H2}) out of the 22 molecules. The reason for this is that updates to specific sets of molecules, whilst not updating others, can change the predicted results (i.e., three sets of cross sections give three results). Whilst this seems obvious, we explicitly say this because it is factual for multiple models used by the community and for results scattered throughout the literature. We found that significant differences (up to a factor of 10 billion) could occur for important molecules such as \ce{NH3} in the Hycean scenarios, although this was reduced from an initially small abundance of $\sim 1$ pptv. Testing every single combination of molecules would require millions of simulations, but we deduced that the main source of the changes in the simulations with only 6 changed molecular cross sections was due to CO.

Because the \textsf{Atmos} and \textsf{Photochem} codes photodissociate at wavelengths where \textsf{VULCAN} only considers photoabsorption (e.g, see CO in Fig.~\ref{All XS figure}), we implement the \textsf{VULCAN} photoabsorption and photodissociation values together when swapping in the photodissociation values used in \textsf{Photochem} for the W24 setup sensitivity tests (see section \ref{Sensitivity to cross sections results}). One can also only swap photodissociation and leave out the photoabsorption. Either way, due to the discrepancies between the data, incident photons are accounted for in different ways, with some either being allowed to propagate through atmospheric layers they should be absorbed by, or with some photons affecting photochemical reactions when they should not be.

Additionally, molecular photolysis can have multiple photochemical reaction channels, with a branching ratio which represents the likelihood of a photochemical reaction channel occurring at a particular wavelength. We have found that the branching ratios between the three models do not always agree, with discrepancies for \ce{H2O}, \ce{NH3}, \ce{CO2}, \ce{CH4}, \ce{O3}, \ce{SO2}, \ce{CH3OH}, \ce{H2CO}, \ce{C2H4}, and \ce{C2H6}. We show the differences for \ce{H2O}, \ce{NH3}, \ce{CO2}, and \ce{CH4} in Fig~\ref{Branching ratio figure}.

To further complicate matters, the photolysis channels assumed are not always the same. For example, the following \ce{NH3} photolysis channels exist:

\begin{align}
    \label{nh3p1}
    \ce{NH3 + $h\nu$ -> NH2 + H} \\
    \label{nh3p2}    
    \ce{NH3 + $h\nu$ -> NH + H + H} \\
    \label{nh3p3}
    \ce{NH3 + $h\nu$ -> NH + H2}.
\end{align}

\textsf{Photochem} uses channels \ref{nh3p1} and \ref{nh3p3}, \textsf{VULCAN} uses channels \ref{nh3p1} and \ref{nh3p2}, but \textsf{Atmos} uses only channel \ref{nh3p1}. For \ce{CH4}, \textsf{Photochem} includes 3 channels, \textsf{VULCAN} includes 4 channels, and \textsf{Atmos} includes 5 channels. In the case of \ce{C2H6}, \textsf{Photochem} includes 1 channel, \textsf{VULCAN} includes 5 channels, and \textsf{Atmos} includes 6 channels. We also found differences in assumed photolysis channels for \ce{S4}, \ce{SO2}, \ce{H2CO}, \ce{CH3}, \ce{CS2}, \ce{C2H2}, \ce{C2H4}, and \ce{C2H6}. Testing all of the differences in cross sections, branching ratios, and photolysis channels is a significant undertaking and our aim is to point out model differences and the possible arising uncertainties.

\section{The assumed stellar spectrum}
\label{Appendix: The assumed stellar spectrum}

\setcounter{table}{0} 
\renewcommand{\thetable}{C\arabic{table}}

\setcounter{figure}{0} 
\renewcommand{\thefigure}{C\arabic{figure}}

\begin{figure*}[t!]
	\centering
	\includegraphics[width=1\textwidth]{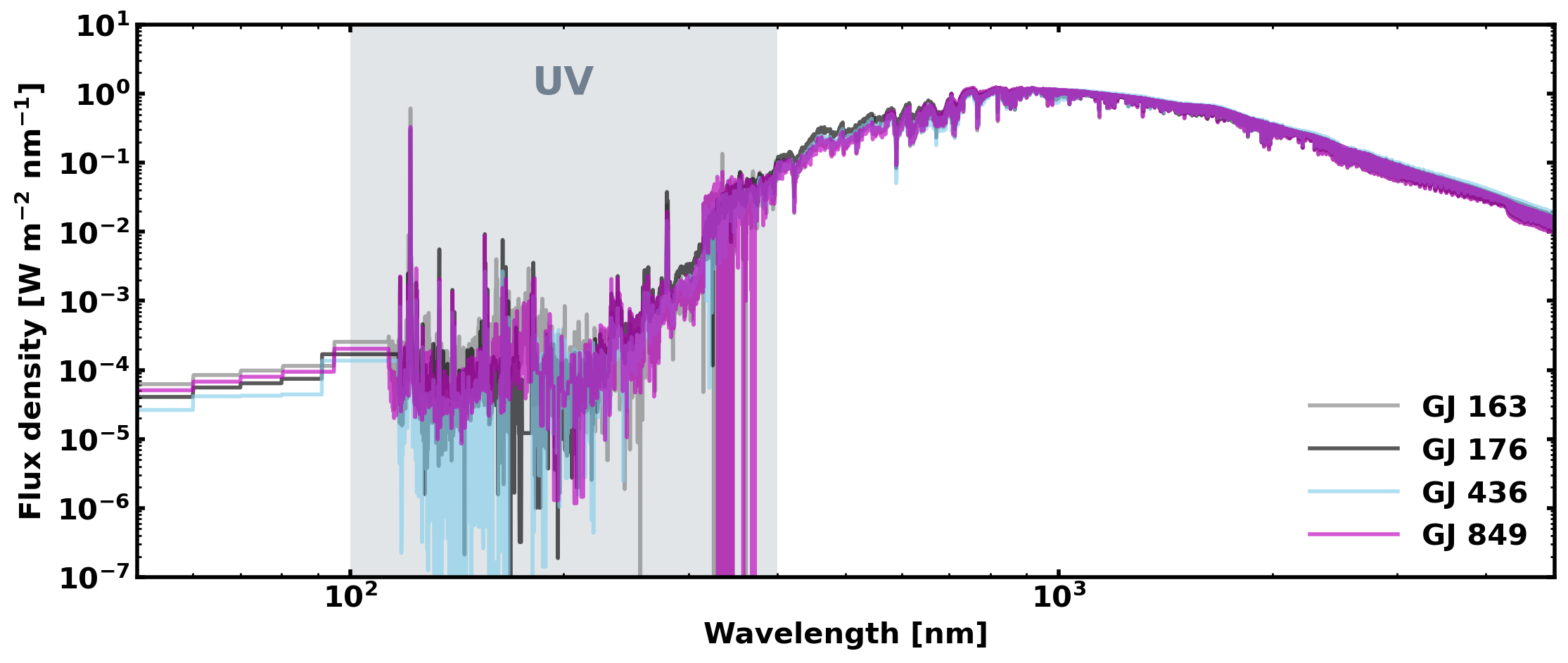}
    \caption{Stellar spectra for four M dwarf stars. The flux density of MUSCLES stars versus wavelength in nm are plotted for the following stars: GJ~163 (grey), GJ~176 (black), GJ~436 (light blue), GJ~849 (magenta). The spectra are normalised to the irradiance K2-18~b receives, which is 1368 W m\textsuperscript{-2}. These stars are all reported to be between M2.5V and M3.5V, with associated errors in observations (see main text, Fig.~\ref{Fractional stellar flux figure}, and Table~\ref{Star spectral type table}).} 
    \label{Stellar spectra figure}
\end{figure*}

\begin{figure*}[t!]
	\centering
	\includegraphics[width=1\textwidth]{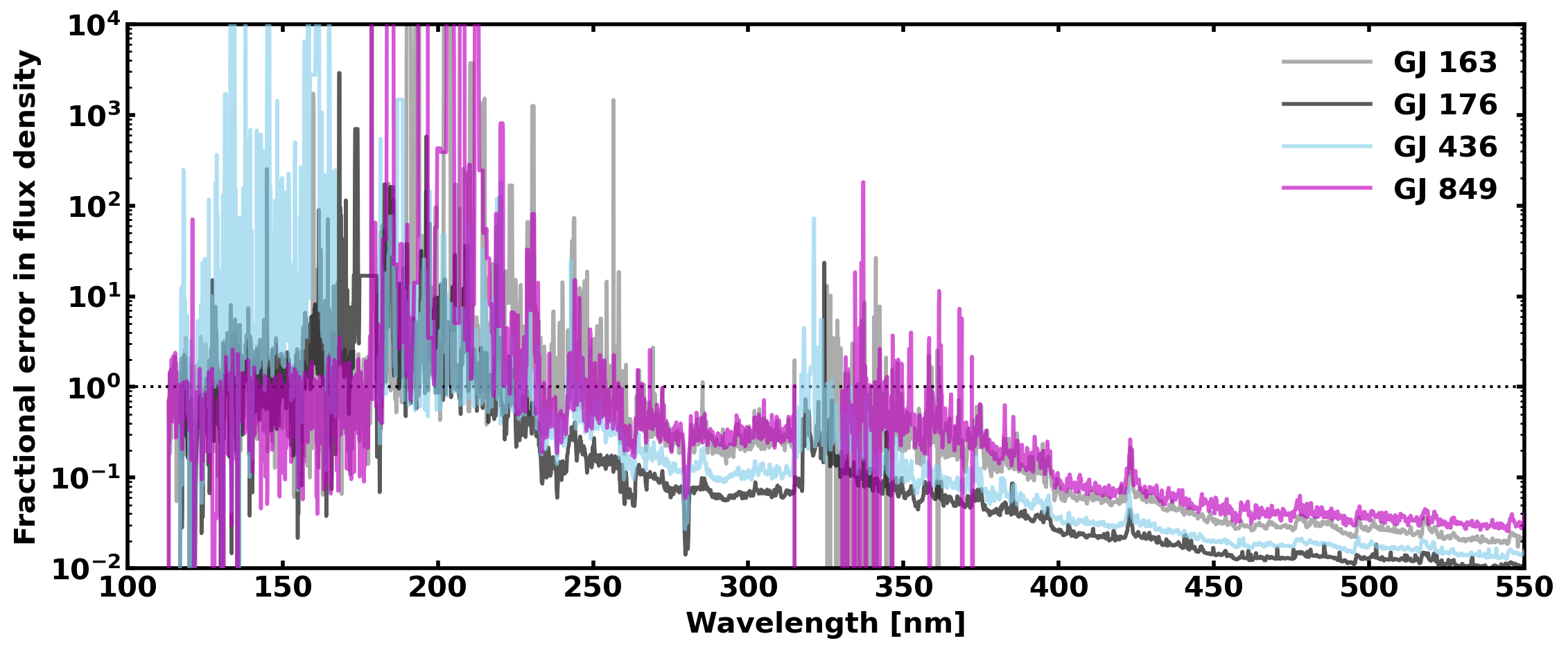}
    \caption{The uncertainty in the stellar spectra for four stars at UV/VIS wavelengths. The fractional error in flux density of MUSCLES stars versus wavelength in nm are plotted for the following stars: GJ~163 (grey), GJ~176 (black), GJ~436 (light blue), GJ~849 (magenta). The black dotted line shows where the reported error in flux is equal to the flux.} 
    \label{Fractional stellar flux figure}
\end{figure*}

When there are no spectral measurements in the UV for a particular star, using another star which closely matches the stellar parameters as an analogue is the most reasonable method for estimating the photochemical environment of the exoplanet's atmosphere. However, the strength and shape of the incoming UV affects molecular photolysis rates, which in turn determines the transmission spectrum that can be retrieved from afar. This is without stellar variability and activity being accounted for, which would further complicate the picture. 

Several MUSCLES stars have spectral types which differ depending on the study. According to \cite{2017ApJ...834..187B}, K2-18 is an M$2.5 \pm 0.5$V star, and \cite{2015ApJ...809...25M} list it as an M2.8V. GJ~176 is listed as an M2.5V star \citep{2014MNRAS.438.2413V, 2009A&A...493..645F} and an M2V star \citep{2007AcA....57..149K}. \cite{2016ApJ...824..102L} lists GJ~436 as a M3.5V dwarf, whilst \cite{2017ApJ...843...31Y} lists it as an M3V dwarf, citing \cite{2012ApJ...753..171V}, who cite \cite{1996AJ....112.2799H} and \cite{1991ApJS...77..417K} for this purpose. \cite{2015A&A...576A..42S} give it as an M2.5V star. TOI-270 is listed as an M3V star by \cite{2023AJ....165...84M} and an M$3.0 \pm 0.5$V star according to \cite{2019NatAs...3.1099G}. Because of the uncertainty on these stellar measurements and their derived spectral type, the spectral types of GJ~849 (M3V), GJ 832 (M1.5V), GJ 674 (M3 V), and GJ~436 (M3.5V) overlap with K2-18, whilst GJ~436 (M3.5V), GJ 674 (M3 V), GJ~163 (M3.5 V), GJ~849 (M3.5 V), GJ 729 (M3.5 V), and GJ~176 (M2.5 V) overlap with TOI-270. Stellar effective temperature ($T_\textrm{eff}$), mass, and rotation can be measured from observations but age constraints are more uncertain \citep{2023AJ....165..195B}. These stellar properties are summarised in Table~\ref{Star spectral type table}.

\begin{table*}[t!]
\centering
\caption{Several different stellar spectra are considered in this study. The spectral type, mass, radius, effective temperature, age, and rotation rate of different stars with associated errors are listed. The relevant references are given here: \textbf{(a)} \cite{2017ApJ...834..187B, 2015ApJ...809...25M, 2018AJ....155..257S, 2019RNAAS...3..189G} \textbf{(b)} \cite{2023AJ....165...84M, 2019NatAs...3.1099G, 2021MNRAS.507.2154V} \textbf{(c)} \cite{2021ApJ...918...40P, 2013A&A...556A.111T, 2015MNRAS.452.2745S} \textbf{(d)} \cite{2014MNRAS.438.2413V, 2009A&A...493..645F, 2007AcA....57..149K, 2021ApJ...918...40P, 2009A&A...493..645F} \textbf{(e)} \cite{2018Natur.553..477B, 2016ApJ...824..102L, 2017ApJ...843...31Y, 2012ApJ...753..171V, 1996AJ....112.2799H, 1991ApJS...77..417K, 2015A&A...576A..42S} \textbf{(f)} \cite{1996AJ....112.2799H, 2007A&A...474..293B, 2019ApJ...871L..26F, 2020ApJ...902....3L, 2023AJ....165..195B, 2021ApJ...918...40P, 2015MNRAS.452.2745S}  \textbf{(g)} \cite{2021ApJ...918...40P, 2020A&A...644A...2I} \textbf{(h)} \cite{2013A&A...549A.109B, 2021ApJ...918...40P, 2023A&A...677A.122P, 2006PASP..118.1685B, 2023AJ....165..195B} \textbf{(i)} \cite{2010A&A...511A..21C, 2021ApJ...918...40P, 2023A&A...678A.207M, 2016ApJ...824..102L}. It is noted that some of the references used list different values for some of the parameters that we list here. This gives added uncertainty to what stellar proxy can be assumed for a particular star and planet.}
\label{Star spectral type table}
\begin{tabular}{@{}ccccccccc@{}}
\toprule
Star & \multicolumn{1}{l}{UV measured?} & Spec type & Mass [$M_\odot$]  & Radius [$R_\odot$] & $T_\textrm{eff}$ [K] & Age [Gyr] & Rotation [d] & References \\ \midrule
K2-18 & No & M$2.5\pm0.5$V & $0.4951\pm0.0043$ & $0.4445\pm0.0148$ & 3457±39 & $2.4 \pm 0.6$ & $39.6\pm0.50$ & a \\
TOI-270 & No & M$3.0 \pm 0.5$V & 0.386±0.008 & 0.378±0.011 & 3506±70 & -- &$\sim 58$ & b \\
GJ~163 & Yes & M$3.5$V & $0.405\pm0.010$ & $0.409^{+0.017}_{-0.016}$ &  $3460^{+76}_{-74}$ & 1 -- 10 &  61.0±0.3 d & c \\
GJ~176 & Yes & M$2$V or M$2.5$V & 0.485±0.012 & 0.474±0.015 & $3632^{+56}_{-58}$ & $8.8^{+2.5}_{-2.8}$ & $40.00\pm0.11$ & d \\
GJ~436 & Yes & M$3\pm0.5$V & $0.445 \pm 0.044$ & 0.449±0.019 & 3479±60 & $8.9^{+2.3}_{-2.1}$ & $44.09 \pm 0.08$ & e \\
GJ 674 & Yes & M2.5V or M3V & $0.353 \pm 0.008$ & $0.361^{+0.011}_{-0.012}$ & $3404^{+57}_{-59}$ & 0.20 & 32.9±0.1 & f\\
GJ 729 & Yes & M$3.5$V & 0.177±0.004 & 0.200±0.008 & $3248^{+68}_{-66}$ & $0.7^{+0.5}_{-0.3}$ & 2.848±0.001 & g \\
GJ~849 & Yes & M$3$V or M$3.5$V & 0.465±0.011 & 0.464±0.018 & $3492^{+70}_{-68}$ & $4.9^{+3.0}_{-2.1}$ & $40.45^{+0.19}_{-0.18}$ & h \\ 
GJ~876 & Yes & M4V or M5V & $0.346\pm0.007$ & $0.372\pm0.004$ & $3201^{+20}_{-19}$ & 0.1 -- 9.9 & $83.7 \pm 2.9$ & i \\
\bottomrule
\end{tabular}
\end{table*}

\cite{2020ApJ...898...44S} used the GJ~176 spectrum in lieu of a well characterised K2-18 UV spectrum, stating that GJ~176 has stellar properties which closely match K2-18. Subsequently, \cite{2021ApJ...921L...8H}, W24, and \cite{2024ApJ...962L...8S} have used the GJ~176 spectrum in place of K2-18. For these two stars, the mass, radius and rotation overlap at the 1-$\sigma$ level, but the stellar effective temperature ($T_\textrm{eff}$) and ages overlap at the 2-$\sigma$ level, though K2-18 is estimated to be much younger ($2.4\pm0.6$ Gyr old) compared to GJ~176 ($8.8^{+2.5}_{-2.8}$ Gyr old). \cite{2024ApJ...966L..24T} recently simulated K2-18~b and varied the stellar spectrum, using GJ~436 (listing it as an M2.5V star) as the closest match to K2-18. GJ~436 is a good match for most parameters in Table~\ref{Star spectral type table}, apart from for the age and rotation rate. \cite{2024ApJ...971L..48Y}, who modelled K2-18~b and TOI-270~d as a mini-Neptune, use the M4V -- M5V star GJ~876 to approximate K2-18. These stars only overlap for age and radius. The mass, $T_\textrm{eff}$ and rotation overlap at $3$-$\sigma$ or worse, such that GJ~876 is not a good proxy for K2-18.

We now consider the properties of GJ~849. For K2-18 and GJ~849, the radius, $T_\textrm{eff}$, and age overlap at the 1-$\sigma$ level, while the rotation and mass overlap at the 2-$\sigma$ level. Both age and rotation rate influence an M dwarf star's UV emission and activity level \citep{2021ApJ...907...91L, 2021ApJ...911..111P}. Hence, one could assume that K2-18 can be approximated by GJ~849 instead of GJ~176 or GJ~436. 

\cite{2024arXiv240303325B} also used GJ~176 in place of TOI-270. These two stars overlap at the 1-$\sigma$ level for $T_\textrm{eff}$, but their masses, radii, and rotation do not overlap at the $3$-$\sigma$ level so these parameters are inconsistent. Again, we consider the properties of a different star. It seems that GJ~163 is the best match to TOI-270 for the stars in Table~\ref{Star spectral type table} when considering mass, $T_\textrm{eff}$, and rotation.

Fig.~\ref{Stellar spectra figure} shows the MUSCLES stellar spectra of GJ~163, GJ~176, GJ~436, and GJ~849, with the UV wavelengths highlighted between 100 -- 400 nm. These spectra are given with uncertainties: the fractional uncertainty in flux density for these stars is shown in Fig.~\ref{Fractional stellar flux figure}. This fractional uncertainty can be very large (up to 10,000) and frequently exceeds 1 for several stars between 100 -- 350 nm. 

\end{document}